\documentclass[a4paper]{llncs}

\usepackage{epsfig}

\newtheorem{thm}{Theorem}
\newtheorem{lem}{Lemma}

\newtheorem{fig}{Figure}

\newtheorem{cor}{Corollary}
\newtheorem{algo}{Algorithm}

\def\leurre{\noindent\leftskip0pt\small\baselineskip 10pt}
\def\grostrait{\ligne{\vrule height 1pt depth 1pt width \hsize}}
\def\demitrait{\ligne{\vrule height 0.5pt depth 0.5pt width \hsize}}

\def\encadre#1#2{%
\setbox100=\hbox{\kern#1{#2}\kern#1}
\dimen100=\ht100 \advance \dimen100 by #1
\dimen101=\dp100 \advance \dimen101 by #1
\setbox100=\hbox{\vrule height \dimen100 depth \dimen101\box100\vrule}
\setbox100=\vbox{\hrule\box100\hrule}
\advance \dimen100 by .4pt \ht100=\dimen100
\advance \dimen101 by .4pt \dp100=\dimen101
\box100
\relax
}

\def\ligne#1{\hbox to \hsize{#1}}
\def\PlacerEn#1 #2 #3 {\rlap{\kern#1\raise#2\hbox{#3}}}

\title{The injectivity of the global function of a cellular automaton in the hyperbolic
plane is undecidable}
\author{Maurice Margenstern\inst{1}}
\institute{%
Laboratoire d'Informatique Th\'eorique et Appliqu\'ee, EA 3097,\\
Universit\'e de Metz, I.U.T. de Metz,\\
D\'epartement d'Informatique,\\
\^Ile du Saulcy,\\
57045 Metz Cedex, France,\\
\email{margens@univ-metz.fr}
}
\begin{document}
\maketitle

\vskip 10pt
\begin{abstract}
In this paper, we look at the following question. We consider cellular automata in
the hyperbolic plane, see \cite{mmJUCSii,mmkmTCS,mmarXivc,mmbook1} and we consider
the global function defined on all possible configurations. Is the injectivity
of this function undecidable? The problem was answered positively in the case
of the Euclidean plane by Jarkko Kari, in 1994, see \cite{jkari94}. In the present paper,
we show that the answer is also positive for the hyperbolic plane: the problem is
undecidable.
\end{abstract}
{\bf Keywords}: cellular automata, hyperbolic plane, tessellations
\vskip 10pt

\def\cqfd{\hbox{\kern 2pt\vrule height 6pt depth 2pt width 8pt\kern 1pt}}
\def\Hii{$I\!\!H^2$}
\def\Hiii{$I\!\!H^3$}
\def\Hiv{$I\!\!H^4$}
\def\norm{\hbox{$\vert\vert$}}
\section{Introduction}

   The global function of a cellular automaton~$A$ is defined in the set of 
all configurations. This is a very different point of view than implementing an
algorithm to solve a given problem. In this latter case, the initial configuration
is usually finite.
  
   In the case of the Euclidean plane, the definition of the set of configurations
is very easy: it is $Q^{Z\!\!Z^2}$, where $Q$~is the set of states of the automaton.

   In the hyperbolic plane, see~\cite{mmbook1,mmhedlund}, following 
what we did in~\cite{mmhedlund}, we have the following situation: we consider that the 
grid is the pentagrid or the ternary heptagrid, see~\cite{mmbook1}. We fix a tile, which 
will be called the {\bf central cell} and, around it, we dispatch $\alpha$~sectors, 
$\alpha\in\{5,7\}$: $\alpha=5$ in the case of the pentagrid, $\alpha=7$ in the case of 
the ternary heptagrid. We assume that the sectors and the central cell cover the plane 
and the sectors do not overlap, neither the central cell, nor other sectors: call them 
the {\bf basic sectors}. Denote by ${\cal F}_\alpha$ the set constituted by the central 
cell and $\alpha$~Fibonacci trees, each one spanning a basic sector. Then, a configuration 
of a cellular automaton~$A$ in the hyperbolic plane can be represented as an element 
of~$Q^{{\cal F}_\alpha}$, where $Q$~is the set of states of~$A$. If $f_A$~denotes 
the {\bf local} transition function of~$A$, its {\bf global} transition function~$G_A$ 
is defined by: $G_A(c)(x)=f(c(x))$.

The injectivity problem for a cellular automaton consists in asking whether there
is an algorithm which, applied to a description of~$f_A$ would indicate whether 
$G_A$~is injective or not.

   In this paper, we prove that there is no such algorithm and so, the corresponding
problem is undecidable. The present paper relies on a previous work by the author,
see~\cite{mmarXivpreinj}. In this paper, we give a construction which is described
in~\cite{mmCSJMprefill,mmarXivprefill}, which yields a plane-filling path, each time 
we can construct a valid tiling with an exception. However, in this exceptional case, 
a more careful analysis of the structure of the path shows that, changing a bit the
way in which triangles and trapezes are traversed by the path, it is also possible to 
carry out the argument which is needed to prove the undecidability of the injectivity.

    Accordingly, we shall not repeat the construction of the interwoven triangles
on which the construction of the mauve triangles rely. In section~2, we
more carefully describe the construction of the path based on the mauve triangles.
In section~3, we show how to derive the proof of the main theorem:

\begin{thm}\label{injundec}
There is no algorithm to decide whether the global transition function of a cellular
automaton on the ternary heptagrid is injective or not.
\end{thm} 

   Note that it is enough to find a particular tiling whose cellular automata have
the property that the injectivity of their global function is undecidable to prove
that the same property for cellular automata in the hyperbolic plane in general is
also undecidable. However, it seems impossible to transfer the construction 
of the path which we consider in this paper to the pentagrid.

\section{A closer look at the mauve triangles}

   Remind that the mauve triangles are first constructed on the interwoven triangles.
The latter triangles are obtained by the following construction, illustrated by
figure~\ref{interwovengreen}.

   We refer the reader to \cite{mmBEATCS,mmCSJMprefill,mmarXive} for 
a detailed account on the construction of the interwoven triangles and for their
properties. We also refer him/her to the same papers for an account on the implementation
of these triangles in the ternary heptagrid of the hyperbolic plane.

   At this point, we would like to make the following remark. 
In~\cite{mmBEATCS,mmCSJMprefill,mmarXive}, we implement the interwoven triangles
in the ternary heptagrid, using another tiling as a background. This tiling, called
the {\bf mantilla}, is a refinement of the ternary heptagrid by grouping its tile
in a particular way. It is possible to implement the interwoven triangles in
a simpler context of the ternary heptagrid. However, the spacing imposed by the mantilla
is a good point which allows to more easily solve a few details of the implementation of 
the path.

   The construction of this tiling needs a lot of signals, which entails a huge number of
tiles, around 18,000 of them, not taking into account the specific tiles devoted to the
simulation of a Turing machine. The construction of this paper requires much more tiles,
but we shall not try to count them.

\vskip 10pt
\setbox110=\hbox{\epsfig{file=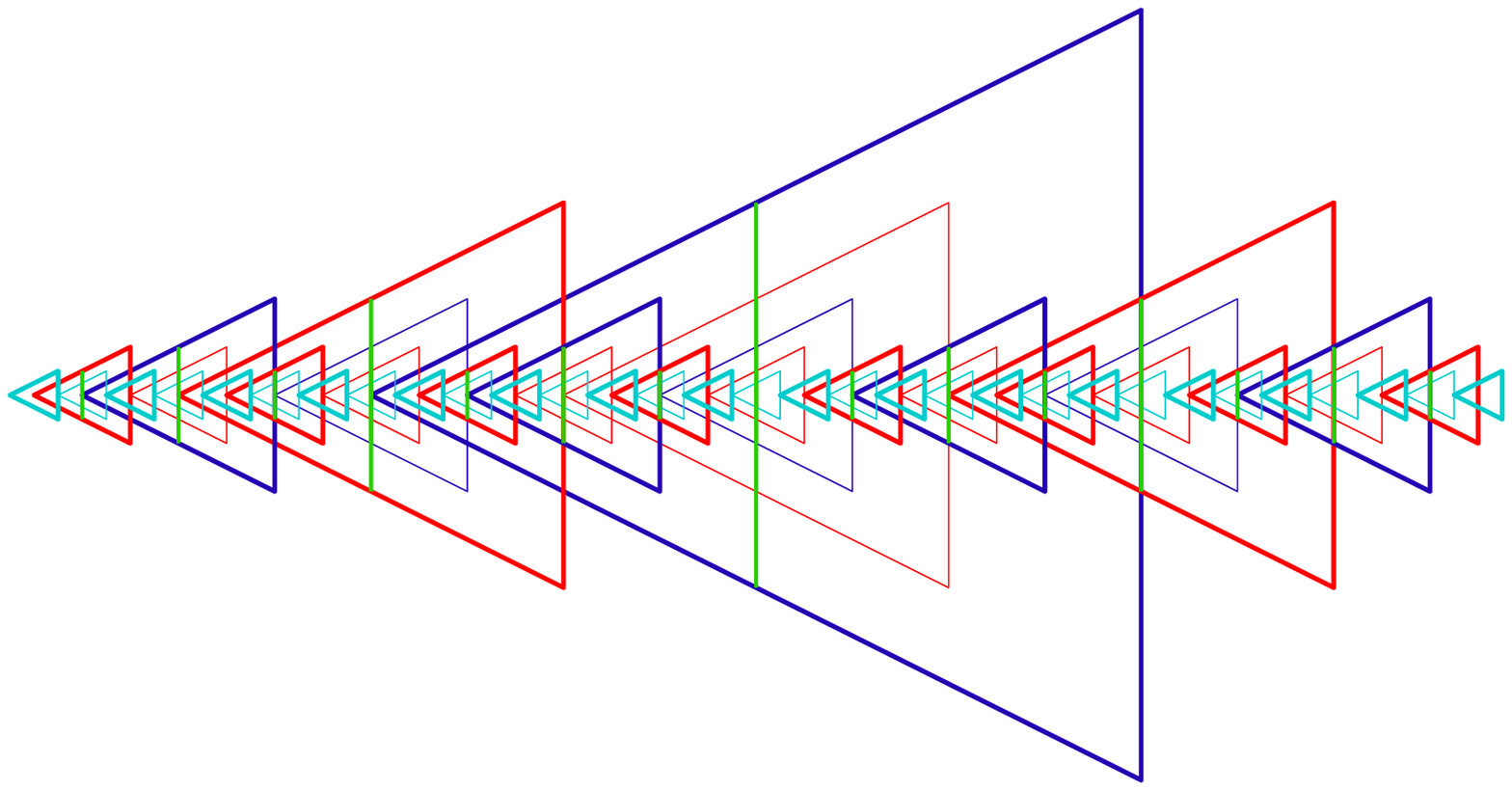,width=400pt}}
\vtop{
\ligne{\hfill
\PlacerEn {-390pt} {0pt} \box110
}
\vskip-45pt
\begin{fig}\label{interwovengreen}
\leurre
Construction of the interwoven triangles in the Euclidean plane: the green signal.
\end{fig}
}
\vskip 10pt


   The mauve triangles are constructed from the interwoven ones. More precisely, we
focus our attention on the red triangles only: they constitute the basis of our
construction. The mauve triangles are simply obtained from the red triangles
as follows. The vertex of a mauve triangle is that of a red triangle~$R$. Its legs 
follow those of~$R$. They go on on the same ray after the corner of~$R$, until they meet 
the basis of the red phantoms of the same generation as~$R$ which are generated by the 
basis of~$R$. At this meeting, the legs meet the basis of the mauve triangle which 
coincide with the basis of the just mentioned red phantoms. 
In~\cite{mmCSJMprefill,mmarXivprefill}, we thoroughly describe the construction 
of the mauve triangles and we refer the reader to these papers. 

   Here, we just mention a few properties of these triangles and then, we shall use
them in order to make it precise the construction of the mauve triangles and the points
attached to them.

\subsection{Properties of the mauve triangles}

   From the above construction of the mauve triangles, we may define the generations 
of the mauve triangles from those of the red triangles: a mauve triangle of the 
generation~$n$ is constructed on a red triangle of the generation~$2n$+1. It will be
later useful to recognize the mauve triangles of generation~0. To this aim, we define
a new colour, called {\bf mauve-0} which is given to these triangles only. Accordingly,
the mauve triangles of generation~0 will be most often called {\bf mauve-0 triangles}.

   From the doubling of the height with respect to the red triangles, the mauve triangles loose the nice property that the red triangles are either embedded or disjoint.
This is no more the case for the mauve triangles. However, we can precisely describe
the overlapping of mauve triangles and how their intersections happen.

   From~\cite{mmCSJMprefill,mmarXivprefill}, we know that the intersection occurs by a leg of a mauve triangle cutting a basis of another mauve triangle. From the construction, any mauve triangle~$T$ of the generation~$n$+1 contains thee mauve triangles of the generation~$n$ with which they have no intersection. They also meet two mauve
triangles of the previous generation. One of them is met at their basis: the legs 
of this triangle of the generation~$n$ cuts the basis of~$T$. The other mauve
triangle~$M$ of the generation~$n$ meets~$T$ near its vertex.
This time, the legs of~$T$ cut the basis of~$M$. We give a number in [0..3] to the
mauve triangles of the generation~$n$ whose vertex is contained in~$T$, as four isoclines are involved by these vertices. Such a number is called the {\bf rank} of the triangles. The rank is periodically repeated on 
\ligne{\hfill}
\vskip-80pt
\vtop{
\setbox110=\hbox{\epsfig{file=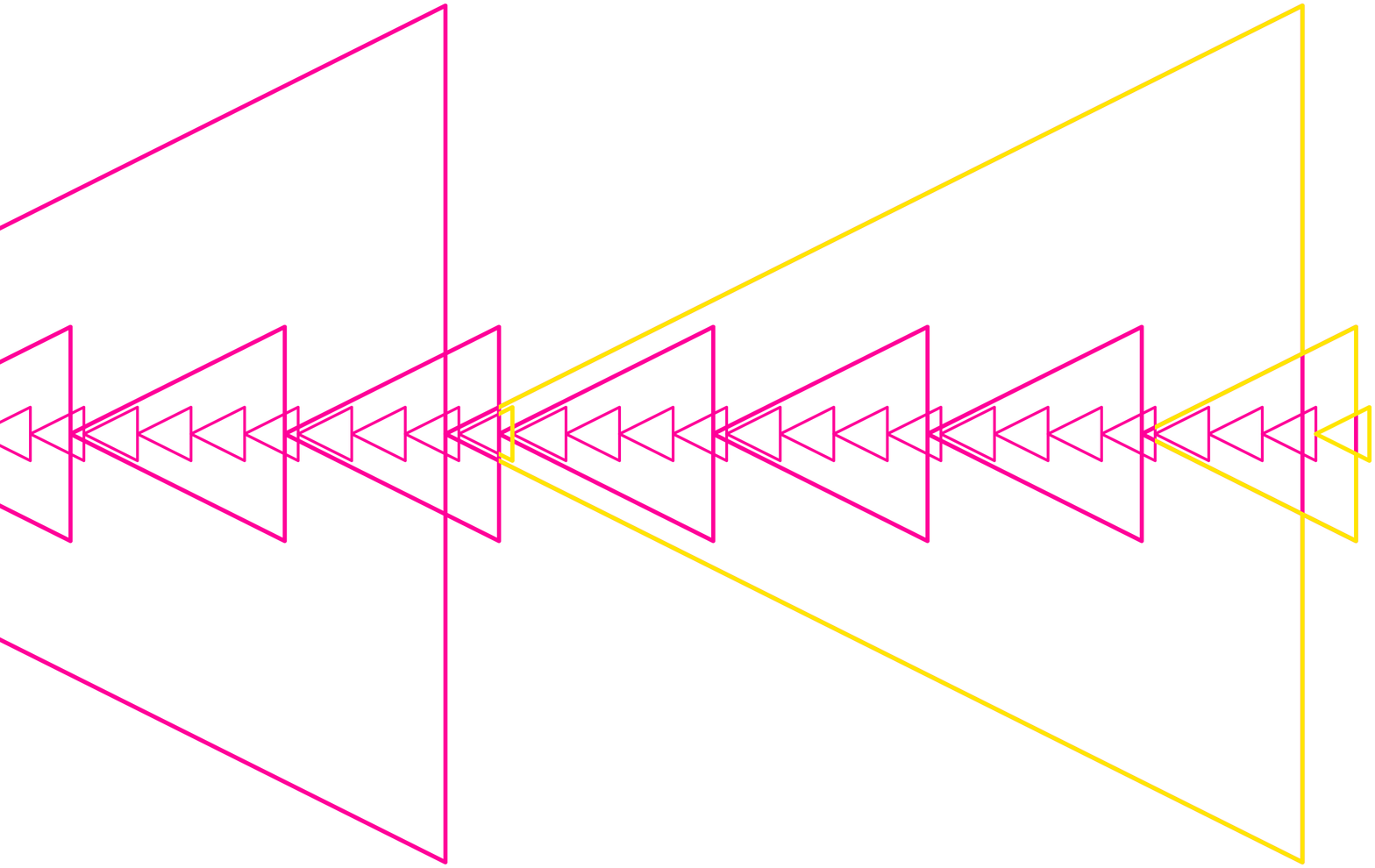,width=250pt}}
\ligne{\hfill
\PlacerEn {-60pt} {0pt} \box110
\hfill}
\vspace{-75pt}
\begin{fig}\label{les_mauves}
\leurre
An illustration of the mauve triangles. 
\end{fig}
}
\vskip -40pt
\noindent
the mauve triangles of the generation~$n$, to the top and to the bottom. A triangle of 
rank~$r$ is called an {\bf $r$-triangle}. If a mauve triangle of the generation~$n$ 
contains~$T$, it is called the {\bf hat} of~$T$: it is a 3-triangle. The hat is unique when
it exists. Note that if we can repeat the construction of the hat recursively until
reaching a mauve-0 triangle, we obtain that the vertex of this mauve-0 triangle
is at a distance at most $\displaystyle{h\over4}$ of the vertex of~$T$. We call the 
mauve-0 triangle the {\bf remotest ancestor} of~$T$, a notion already remarked for 
the interwoven triangles.
Accordingly, if the vertex of~$T$ is on the basis of a mauve-triangle of the same 
generation, then its remotest ancestor exists.

The triangles of the generation~$n$ which cut the basis of~$T$ are also 3-triangles. We 
define the {\bf low points} of a leg of a triangle, $LP$ for short, as the point which 
is at a distance~$\displaystyle{h\over4}$ from the basis of the triangle, where $h$~is 
the length of the leg. The $LP$'s play an important role: the line which joins 
the~$LP$'s of~$T$ cuts the 2-triangles also at their $LP$'s. The intersection of the 
basis of~$T$ with its 3-triangles occur at their $LP$'s. 

In~\cite{mmCSJMprefill,mmarXivprefill}, the consideration of the $r$-triangles has 
led to the extension of the notion of latitude used in the interwoven triangles to the 
case of the mauve triangles. However, the definition which was there given is not 
satisfactory. In order to introduce a better one, we define the {\bf primary latitude} 
of a mauve triangle as the set of isoclines which cross its legs, the basis being 
included but the top being excluded. This allows us to obtain a partition of the hyperbolic 
plane by the primary latitudes attached to a given generation. This defines a 
partition for each generation. But the primary latitudes may overlap from one generation 
to the next one.

   Now, we can precisely state the properties mentioned above about the intersections between mauve triangles and we can also prove them as follows.

\begin{lem}\label{innerandLP}
Let $T$ be a mauve triangle of the generation~$n$$+$$1$. Then, the primary latitude
of~$T$ intersects five primary latitudes of the generation~$n$, denote them
by~$L_{-1}$, $L_0$, $L_1$, $L_2$ and $L_3$. There are four triangles of the generation~$n$,
$T_0$, $T_1$, $T_2$ and $T_3$ with the following properties:
\vskip 3pt
{\leftskip 20pt\parindent 0pt
$(i)$  $T_i$~belongs to the primary latitude~$L_i$ for~$i$ in~$\{0,..3\}$;
\vskip 0pt
$(ii)$ the vertex of~$T_{i+1}$ is on the basis of~$T_i$ for $i$~in $\{0,..2\}$;
\vskip 0pt
$(iii)$ the $LP$'s of~$T$ and~$T_2$ are on the same isocline;
\vskip 0pt
$(iv)$ the legs of~$T_3$ cut the basis of~$T$ at their $LP$'s.
\par}
\vskip 3pt
\noindent
When it exists, $T_{-1}$ contains the vertex of~$T$ and it belongs to the latitude~$L_{-1}$.
In this case, the vertex of~$T$ is on the isocline which joins the $LP$'s of~$T_{-1}$.
\end{lem}

    Let~$R$ be the red triangle whose vertex and legs support those of~$T$.
Its generation can be written as~$2m$+3. It contains two red latitudes of the 
generation~$2m$+1, each one containing a triangle~$R_\alpha$, with $\alpha\in\{1,2\}$,
and we may assume that $R_2$~is below~$R_1$ and that the vertex of~$R_2$ belongs to the
basis of a red phantom whose vertex belongs to the basis of~$R_1$. 

From \cite{mmBEATCS}, we know that $R$~contains several red triangles of the 
generation~$2m$+1 which belong to~$L_1$. We pick one of them: it is~$R_1$. It generates 
a mauve triangle~$T_0$ which is contained in~$T$, as $R_1$~is also contained in~$T$. 
Note that $T_1$~is generated by a red triangle of the generation~$2m$+1 which
is on the second latitude of this kind which are contained in the latitude of~$R$.
Now, the second half of the legs occurs in the latitude of red phantoms of the 
generation~$2m$+3. We know that inside red phantoms, the structure of the inner trilaterals
is the same as inside a red triangle of the same generation. This defines~$R_2$ 
and~$R_3$ which are red triangles of the generation~$2m$+1. They are inside~$T$ and
we have the same relation between~$R_1$ and~$R_2$ 
as between~$R_2$ and~$R_3$. And so, we find again two
red triangles which give rise to~$T_2$ and~$T_3$. The succession 
from~$T_0$ to~$T_1$ and then from~$T_1$ to~$T_2$ and, finally, from~$T_2$
to~$T_3$ is the same. By the situation of~$R_4$, we obtain that the legs of~$T_3$ cut
the basis of~$T$ at their $LP$'s. We also see that the basis of~$T_2$ is on the isocline
of the vertex of~$T_3$. As the distance from the basis of~$R_3$ to the mid-distance line 
of the phantom~$Q$ of the generation~$2m$+3 which contains~$R_2$ and~$R_3$, is the same
as the distance from the mid-distance line of~$Q$ to the vertex of~$R_4$, and that this 
distance is half the height of a trilateral of the generation~$2m$+1, we have that
the mid-distance line of~$Q$ is both on the isocline of the $LP$'s of~$T$ and also
on the isocline of the $LP$'s of~$T_2$. This proves the points~$(iii)$ and~$(iv)$.
The proof of the other points is also contained in the just given arguments. 

   Now, assume that there is a red triangle~$R_0$ of the generation~$2m$+1 which is in 
the same connection with~$R_1$ as $R_1$~is with respect to~$R_2$. This means that
there is a blue triangle~$B$ of the generation~$2m$+2 which generates the vertex of~$R$
on its mid-distance line and the vertex of~$B$ is generated by~$R_0$.
Now, $R_0$~generates a mauve triangle~$T_{-1}$ which contains the
vertex of~$T$ as there is a red phantom~$P$ in the tower of phantoms around the
vertex of~$R$, whose vertex is on the basis of~$R_0$ and whose basis generates the
vertex of~$R_1$. Accordingly, $T_{-1}$~contains the vertex of~$T$ as~$T_{-1}$ also
contains~$P$. 

   Now, note that the vertex of~$T$ is on the mid-distance line of the phantom~$P$.
As the height of the triangle~$T_{-1}$ is twice that of~$R_0$, we have that $T$~is
on the isocline which supports the mid-distance line of~$P$ and, according to the just 
performed estimation, this isocline joins the $LP$'s of~$T_{-1}$. This 
proves the last assertion of the lemma.
\cqfd

   From the proof, we can deduce a way to determine the mid-point and the $LP$'s
of a mauve triangle by means of a finitely generated signals. It will be enough to show that we have just to append a few signals to those which we
have already defined.

\subsection{Construction of the $LP$'s of a mauve triangle}

   Let $T$ be a mauve triangle of the generation~$n$ and let~$h$ be its height. Its 
mid-points, which lay at a distance~$\displaystyle{h\over2}$ are easy to determine: it 
is the corner of the red triangle~$R$ on which $T$~is constructed. 

    The determination of the $LP$'s needs some work. In fact, the $LP$'s of~$T$ are 
on the same isocline as the mid-distance line of a phantom whose vertex is generated by 
the basis of~$R$. Let $P$~be the leftmost phantom generated by
the basis of~$R$. As $P$ is a phantom, its green signal may cut the leg of~$P$. 
However, it may not as well, see \cite{mmBEATCS}, in which case it is continued by 
an orange signal. Now, the leg of~$T$ meets a lot of orange signals coming from all 
the phantoms which stand close to its borders. Without further indication, the leg 
cannot distinguish which one comes from~$P$. 

   The construction proceeds as follows:

   First, we look at the determination of the corners of~$T$ itself which is not 
that straightforward.

 At the corners of~$R$, the 
mauve signal defining the leg of~$T$ goes on along the extremal branch of the 
Fibonacci tree defining~$R$. At the same time, each corner of~$R$ sends a signal 
towards the other one on the isocline of the basis of~$R$. Call this signal 
the {\bf brown} signal. The brown signal has the laterality of the corner. When the 
brown signal meets the first vertex of the phantom~$P$, it is a red phantom of the 
generation~$2n$+1. The signal goes down along the leg of the phantom which has its 
laterality. It goes along this leg until it meets the corner of~$P$. There, on the 
isocline~$\iota$ of the basis of~$P$, the brown signal leaves the leg to run 
on~$\iota$, to the side of its laterality, until it meets
the mauve signal of the leg of~$T$. Then, a mauve signal is sent to the other side, in 
order to meet the mauve signal sent by the other corner of~$T$. 

Now, the problem for the signal is to meet the correct leg, as it may encounter a 
lot of them along~$\iota$, the isocline of the basis of~$P$, which belong to smaller 
generations. The signal may cross them because it meets them at their $LP$'s where 
this possibility is foreseen. When the brown signal first meets a leg whose laterality 
is opposite to its one, it knows that it is not the appropriate leg.
Now, the signal does not cross the leg: otherwise it will meet the opposite leg which
is also not the right one: it might also meet several legs of its laterality before 
meeting a leg with an opposite laterality. As the brown signal cannot count arbitrary 
numbers, it circumvents the triangle by climbing along its leg up to the vertex and 
then going down to the appropriate isocline. In order
to recognize the right isocline, when the brown signal starts its circumventing path,
it sends another brown signal, say a light one, with no laterality, which goes on
running on~$\iota$, towards the appropriate corner. As $\iota$~is an isocline of
the $LP$~of the triangle which the brown signal circumvents, the brown 
signal cannot meet another light brown signal meeting the leg: as for red trilaterals, the
isocline of a basis is specific to any mauve triangle. And so, when going down along
the leg, the brown signal meets its light brown one, it knows that it has found~$\iota$
on which it goes on its way, still to the side of its laterality. And now, the first
mauve leg of its laterality met by the brown signal is the right one.

   Now, we can use the brown signal in order to locate the~$LP$'s of~$T$. Consider
the time when the previous brown signal is going down along the appropriate leg of~$P$.
When the brown signal meets the mid-point of~$P$, it knows that it is the isocline of
the $LP$'s of~$T$. And so, the brown signal sends a purple signal of the same
laterality as the brown signal towards the side of its laterality on the 
isocline~$\zeta$ of the mid-distance line of~$P$. This signal also circumvents the mauve 
triangles which it meets. Now, the signal is able to recognize~$\zeta$ during the 
circumvention of phantoms thanks to the following. We know that the purple signal
meets smaller mauve triangles at their $LP$'s. By induction, we assume that
a similar signal arrives to the $LP$'s from inside the mauve triangle~$M$, created at
the time of the construction of~$M$. Note that in any case, such a signal is stopped
by the leg of~$M$. Now, the arriving signal from the mid-point of the leg of~$P$
is deviated to the first part of the leg of~$M$. When the signal goes down on the
other leg, it identifies its~$LP$ by the arrival of a similar signal of the appropriate
laterality which is stopped by the leg. This allows the signal to again find~$\zeta$ and
to go on its route on this isocline. Due to the laterality of the purple signal and to
the fact that its laterality is unchanged and that it must match the mauve leg it meets
from inside, such a signal cannot be present if it is not sent by a brown signal for
detection purpose.

   Let us look closer at this assertion.

   Consider a mauve triangle~$M_0$ and, inside it, a mauve triangle~$M_1$ in a place 
where the isocline of the $LP$'s of~$M_1$ do not meet those of~$M_0$. Assume that
the generation of~$M_0$ is the successor of the generation of~$M_1$. Assume too that
an inner signal~$\sigma$ of~$M_1$ goes out until the leg of~$M_0$. If this happens, this may
confuse a signal which is circumventing~$M_0$. Now, as~$\sigma$ goes out from the 
$LP$ of~$M_1$, $\sigma$~also circumvent~$M_1$ and, together, all the other
maximal inner mauve triangles of a generation not greater than that of~$M_1$ whose $LP$'s 
are on the same isocline as those of~$M_1$. Consequently, $\sigma$ reaches both
legs of~$M_0$. Now, $\sigma$ has a single laterality and when it meets a mauve leg from
inside, its laterality must be the same as that of the leg. But here, this not possible,
and so $\sigma$ cannot go out of~$M_1$.
   
 And so, this ensures the detection of the $LP$'s. Note that the
detection of the $LP$'s of inner mauve triangles with the same isocline~$\zeta$
does not alter the process due to the fact that the leg stops the signal. There are tiles
with a circumventing signal, with a laterality independent on that of the leg, and tiles
without it but, in both cases, the finishing signal inside the triangle is present with the
appropriate laterality. These two kinds of tiles, and no others, force the right choice,
again as laterality cannot be changed. 
\vskip 5pt
   Now, the mauve signal which defines the basis of a mauve triangle is also emitted
by the vertices of the mauve triangle of the same generation but whose primary 
laterality is just below the considered one. Now, in mauve triangles, a basis must be 
stopped by its corners.

   To handle this situation, we remark that the constraint of laterality of the 
brown signal allows us to forbid the existence of such a signal in between two 
contiguous red phantoms of the generation~$n$ with their vertices on the basis 
of~$R$. Now, if a mauve triangle is missing, the left-hand side brown signal goes 
to the left until it meets the corner of a mauve triangle which exists within the 
considered latitude. Now, such a meeting from outside of the corner is ruled out: 
it is enough to forbid such a configuration for the tile defining the corner. And 
so, if there is no red triangle, the brown signal will be destroyed, as the 
phantom without brown signal does exist within the same latitude.  For the mauve 
basis, it is an analogous situation: it may be emitted by the vertex of a mauve 
triangle and it may be not. If it is emitted and if there is no leg to stop it, 
the basis meets a corner of a mauve triangle from outside. As the corner stops the
mauve basis which is inside the triangle, it is enough to not allow any merging at the
level of this tile. And so, if the mauve legs do not go down, the basis cannot
exist: it is ruled out by external corners. Now, if the vertex exists, the legs exist and
also the corners, due to the brown signal and the corners force the presence of a 
unilateral mauve basis in between them.

   We can now state:

\begin{lem}\label{tileLP}
The mauve triangles together with the determination of their $LP$'s and 
mid-points can be constructed from a finite set of prototiles.
\end{lem}
\vskip 5pt

   Later, inside a triangle or a zone in between two triangles of the same primary
latitude, the path which we shall later describe will completely cross the primary
latitudes of these domains. This is why it is important to clearly delimit each zone 
in each kind of area. For this purpose, we introduce the notions of 
$\beta$-cline which we study in the next subsection, together with its construction.

\subsection{The $\beta$-clines and their construction}

   We start from the remark that the basis of a mauve triangle~$T$ of the 
generation~$n$+1 cuts the legs of the 3-triangles which have their vertex 
inside~$T$. Repeating this remark to the 3-triangle of the generation~$n$,
we can construct a sequence $\{T_i\}_{i\in [0..n+]}$ such that:
\vskip 5pt
{\leftskip 20pt\parindent 0pt
$(i)$ $T_{n+1} = T$;

$(ii)$ $T_i$ is a 3-triangle of the generation~$i$ for~$i$ in $[0..n]$;

$(iii)$ the basis of $T_{i+1}$ cuts the legs of~$T_i$, of course at their $LP$.
\par}

   Any mauve triangle~$T$ of a positive generation generates such a sequence which 
we call the {\bf shadow} of~$T$ and that $n$+1 is its generation. Of course, 
if $\{T_i\}_{i\in [0..n+1]}$ is the shadow of~$T_{n+1}$, the sequence 
$\{T_j\}_{j\in [0..i+1]}$ is the shadow of~$T_{i+1}$ for~$i$ in $[0..n]$. We say that 
the shadow $\{T_j\}_{j\in [0..i+1]}$ is a {\bf trace} of the 
shadow $\{T_i\}_{i\in [0..n+1]}$.

   We say that a shadow $\{T_i\}_{i\in [0..n+1]}$ is a {\bf tower}, if it is finite  
and if it is not the trace of a shadow of a bigger generation.
We shall see that there may be a sequence of mauve triangles 
$\{T_i\}_{i\in I\!\!\!N}$ in which $\{T_i\}_{i\in [0..n+1]}$ is a trace of
$\{T_i\}_{i\in [0..n+2]}$ for any~$n$. In this case, we say 
that $\{T_i\}_{i\in I\!\!\!N}$ is an {\bf infinite tower}.

    When $\{T_i\}_{i\in [0..n+1]}$ is a finite tower, we say that the isocline 
of the basis of~$T_0$ is the {\bf $\beta$-cline} of~$T_{n+1}$ and that 
its {\bf type} is the rank of~$T_{n+1}$. 

   From the $\beta$-clines, we define two new points on the legs of a triangle 
of a positive generation: the~$\beta$- and~{\bf $\gamma$-points}.

   By definition, the $\beta$-point of a mauve triangle~$T$ of the 
generation~$n$+1 is the intersection of its leg with the $\beta$-cline of 
the 2-triangles whose vertex in inside~$T$. It is not difficult to see that 
the $\beta$-point lies on the leg in between the $LP$ and the corner. It is 
at a distance less than $\displaystyle{h\over{12}}$ from the line joining 
the $LP$'s of~$T$, with $h$~being the height of~$T$, and as closer to this value 
as $n$~tends to infinity.

\subsubsection{Constructing the $\beta$-cline}

   To construct the $\beta$-cline, we define signals which start from the $LP$'s
of a mauve triangle of the considered generation and latitude. Call them the
\hbox{\bf $\beta$-signals}. The $\beta$-signals are lateral, with the laterality which is 
opposite to that of the leg on which they start. They travel along legs of mauve 
triangles and along isoclines of a basis. The $\beta$-signals go down along legs of 
a laterality opposite to their own one, from an $LP$ to a corner. When they run
on an isocline, they go in the direction of their laterality. When they meet a 
corner, they run on the basis, in the direction of the other corner. They can freely 
travel on this isocline, until they meet the leg of a triangle of a laterality which 
is opposite to their own one and at their $LP$. If the leg is of another laterality
or if the meeting point is not in the closed interval with the $LP$ of the leg and 
its corner as end points, the $\beta$-signals crosses the leg. It is plain that 
both $\beta$-signals starting from the opposite $LP$'s of the same mauve triangle 
will meet, and they cannot do that along a leg or at a corner. 
When they meet, they use a join tile, see~\cite{mmBEATCS}, in which the right-hand, 
left-hand side $\beta$-signal is on the left-, right-hand side part of the tile. It is 
plain that both $\beta$-signals define a kind of convex hull of this part of the mauve 
triangle. 

   Note that for two consecutive mauve triangles of the same generation within the same
isocline, the $\beta$-signal which starts from the low-point of one of them
cannot travel on this isocline to the facing low-point of the other triangle.
Indeed, on the right-hand side low-point, we have a left-hand side $\beta$-signal and
on the left-hand side low-point with have a right-hand side $\beta$-signal. And so,
this would require a join tile with a left-hand side $\beta$-signal on the left-hand 
side part of the tile: this is ruled out.

   We may impose an additional constraint on the join tile for $\beta$-signals
of opposite lateralities with the right-hand side signal on the left-hand side 
of the join tile: the join tile generates a horizontal unilateral yellow signal. 
This signal runs on an isocline only: it marks the $\beta$-cline. It is important
to note here that the whole isocline constitutes the $\beta$-cline.

   By construction, the yellow signal travels along an isocline of a basis of a mauve~0 
triangle. Consequently, it travels on an isocline~5. Accordingly, it meets no 
basis of a mauve triangle of a positive generation and no $LP$, as $LP$'s are always on 
an isocline~15. And so, the yellow signal will meet legs of triangles. 

\ifnum 1=0 {  
   Let us closer examine which legs can be met by the yellow signal and how to codify
the resulting information on the signal.

   Consider a yellow signal produced by a mauve triangle~$M$ of a positive generation~$n$.
If $M$~is inside a mauve triangle~$T$ of the generation~$n$+1, the yellow signal will
meet both legs of~$T$ from inside. In this case, we do nothing.

   If $M$~is not inside a triangle of the generation~$n$+1, it may be in between two
such triangles~$T_1$ and~$T_2$ which are consecutive and within the same latitude
with, for instance, $T_1$ on the left-hand side of~$M$ and~$T_2$ on its right-hand side.
Then, the yellow signal will meet the right-hand side leg of~$T_1$ and the left-hand side
leg of~$T_2$. If it meets the part of one leg covered by the $\beta$-signal, the meeting
of the other leg also happens on a part covered by the $\beta$-signal. Then, this meeting
rises a black signal which accompanies the yellow one.

   If the yellow signal meets the leg of~$T_1$ on another part of the leg, we do nothing.

   The last situation, is when $M$~is outside a triangle of the generation~$n$+1 and within
its latitude and also inside a triangle of a generation~$n$+$k$+2, with $k\geq0$.

   Then we decide that the decision relatively to a possible black signal is given
by the triangle of the generation~$n$+1. If the black signal is present, it is stopped
by the leg of the triangle of the generation~$n$+$k$+2.

   Accordingly, a wrongly placed tile can be removed without ambiguity and it can be
replaced by the appropriate tile with, or without the black signal as, when the yellow 
signal reaches a leg from outside, the leg requires the black signal or forbids it.
Now, on the other side of the yellow signal, either the same situation occurs, or we have
the neutral situation of a leg reached from inside which allows both situations. In this
latter case, the situation of the leg reached from outside determines the signal.
}
\fi   

   In our study of the shadow of a mauve triangle, we have already noticed that the 
same $\beta$-cline can be shared by several triangles of different generations.

For the purpose of the path, in the case of a $\beta$-cline of type~2, we consider
that it defines a special signal on the isocline which is just below the $\beta$-cline.
This means that there is a {\bf pre-path signal} on the isocline~4 which is just below
a $\beta$-cline of type~2. This signal plays an important role as can be seen further.
Note that the pre-path signals of a given generation are different from those of
the next generation: the signals corresponding to the generation~$n$+1 occur on the 
isocline~4 of a $\beta$-cline of type~3{} in terms of the generation~$n$. We have a 
stronger result:

\begin{lem}\label{isoprepath}
The isoclines of a pre-path signal of the generation~$n$ are different from those of the
generation~$m$ for any $n$, $m$ with $n\not= m$.
\end{lem}

\noindent
Proof.
Consider the smaller generation, say~$m$. The pre-path signals of the next generation 
correspond
to $\beta$-clines of type~0 of the generation~$m$. Now, for the generation~$m$+2 we have 
the same correspondence with the $\beta$-clines of type~3 of the generation~$m$+1. 
But,
the $\beta$-clines of type~3 of a generation also belong to the $\beta$-clines of type~3 too
of the previous generation. This is due to the fact that the number of $\beta$-clines which
are crossed is a multiple of~4. By induction, this proves our claim. \cqfd


   

\subsection{The $\beta$-points and their construction}

   For the next section , we need to make clear the connection between a mauve 
triangle~$T$ of the generation~$n$+1 and its inner mauve triangles of the generation~$n$.

   To locate the triangles of the just previous generation, there is a way given by the
local numbering of the triangles. We already noticed that the intersection between 
mauve triangles occur between a leg and a basis and that with respect to the leg, the 
intersection happens at its low point. The consequence is that mauve triangles 
of the generation~$n$ which are inside~$T$ are cut by the basis of~$T$ if and only if
they are 3-triangles. Now, the converse is true:

\begin{lem}\label{cutis23}
Let $T$ be a mauve triangle of the generation~$n$$+$$1$. Its basis cuts mauve triangles
of the generations~$i$ for any~$i$ in~$[0..n]$. When $i=n$, the mauve triangle is of 
type~$3$. When $i<n$, the mauve triangle is of type~$2$. 
\end{lem}

\noindent
Proof.
We already know the property for $i=n$. If $n=0$ we are done. If not, consider $i=n$$-$1
and let $T_1$ be a mauve triangle of the generation~$n$ and of type~3 which meets the
basis of~$T$. It is not difficult to see, by definition of the $LP$'s of~$T_1$, that
the line of the $LP$'s of~$T_1$ cuts inner mauve triangles of the generation~$n$$-$1
if and only if they are of type~2. Now, this can be repeated for these 2-triangles. Now,
as mauve triangles which are within the same latitude have the same type, we conclude
that any mauve triangle of the generation~$i$, with $i<n$, is of type~2. \cqfd

   Now, consider the $\beta$-cline of type 2 which corresponds to the mauve triangles
of the generation~$n$ which are inside~$T$. This $_beta$-cline gives rise to a pre-path
which plays an important role in the construction of a path which fills~$T$. In particular,
it is important to recognize the intersection of this $\beta$-cline with the legs of~$T$.
We call them the {\bf$\beta$-points} of~$T$.

   Lemma~\ref{cutis23} gives us a way to construct the $\beta$-points of a mauve triangle.

   Note that there is no $\beta$-point on a mauve-0 triangle and that the $\beta$-point
of a mauve triangle~$T_1$ of generation~1 is easy to determine. Indeed, the line joining the
$LP$ of~$T_1$ cuts inner mauve-0 triangles of type~2. Now, the basis of theses
triangles are on the same isocline which is the $\beta$-cline passing through the 
$\beta$-point of~$T_1$. And so the construction is simple:
\vskip 5pt
\vtop{
\grostrait
\vskip 3pt
{\leftskip 40pt\rightskip 20pt\parindent -20pt
a {\bf silver} signal is sent from the~$LP$ of~$T_1$ until it reaches the 
first mauve-0 triangle of type~2, $T_2$;

the silver signal goes down along the leg of~$T_2$; when it reaches the corner of~$T_2$,
it also reaches the $\beta$-cline of~$T_2$; it follows this $\beta$-cline outside $T_2$;

this intersection of the silver signal with the leg of~$T_1$ defines the $\beta$-point 
of~$T_1$. 
\par}
\demitrait
}
\vskip 5pt
   The construction of the $\beta$-point in the general case relies on lemma~\ref{cutis23}.

   Now, this time, we cannot start from the~$LP$ of~$T$: it meets only triangles
of type~2 of all generations from~0 to~$n$ and so, it is difficult to recognize those
of generation~$n$ by a fixed in advance amount of means.

   Now, if we start the silver signal from the corner, it can be performed according to the
following algorithm:

\vtop{
\begin{algo}\label{betapoint}
\leurre
The construction of the $\beta$-point of a triangle~$T$ of the generation~$n$$+$$1$.
\end{algo}
\vspace{-12pt}
\grostrait
\vskip 3pt
{\leftskip 40pt\rightskip 20pt\parindent -20pt

the silver signal starts from the corner into two directions;

the first direction follows the basis until it meets the leg of a triangle of type~$3$;

\leftskip 60pt
it goes along this leg up to the vertex and there, it follows the isocline of the vertex
away from the leg of~$T$, until it meets a corner which is a corner of a triangle 
of type~2, $T_2$;

from the corner of~$T_2$, the silver signal follows the $\beta$-signal coming from 
the~$LP$ of~$T_2$ which is above the considered corner;

then the silver signal eventually meets the $\beta$-cline defined by the $\beta$-signal 
of~$T_2$; the silver signal goes back to the leg of~$T$, following the just met 
$\beta$-cline;

\leftskip 40pt
the second direction follows the leg of~$T$ reaching the corner and goes up along this
leg towards the~$LP$ of~$T$;

both directions of the silver signal meet at the intersection of the leg of~$T$ with the
expected $\beta$-cline of type~2 coming from an internal mauve triangle of the 
generation~$n$: it is the expected $\beta$-point and the intersection
stops both silver signals.
\par}
\demitrait
}
\vskip 5pt
   Note that this construction also holds when $n=0$.

   Let us prove that algorithm~\ref{betapoint} is correct.

   From lemma~\ref{cutis23}, when the first direction of the silver signal meets a 
3-triangle, it is a triangle~$T_3$ of the generation~$n$. In fact, the signal meets the 
leftmost 3-triangle of the generation~$n$ whose vertex is inside~$T$. When the signal 
arrives on the isocline of the vertex of~$T_3$, it is the isocline of the bases of the 
2-triangles of the generation~$n$ which are inside~$T$. Going away from the initial leg 
of~$T$, the signal arrives to the 2-triangle of the generation~$n$ which is the closest 
to the leg, $T_2$. The signal meets $T_2$ at its corner which faces the leg of~$T$. At 
this corner, according to what we have seen, the silver signal meets the $\beta$-signal. 
As it then follows the $\beta$-signal down to the $\beta$-cline of~$T_2$, and then 
the $\beta$-cline itself in the direction of the leg of~$T$ from where it comes, it 
eventually meets the leg of~$T$. Now, the silver signal meets legs of both lateralities
which cannot always been distinguished even with the help of the circumventing technique. 
And so, the first direction of the silver signal goes on along the $\beta$-cline, constantly.
Now, during the same time, the second direction of the signal goes up along the same leg 
of~$T$ from the corner to the direction of the~$LP$. Before reaching the~$LP$,
it meets the $\beta$-cline which is accompanied by the other silver signal. This allows
to fix the $\beta$-point and both directions of the silver signal to meet and to stop 
each other. \cqfd 

   Note that these silver signals of different mauve triangles can meet: this happens
when the first direction of the silver signal coming from~$T$ reaches the corner of 
the 2-triangle of the generation~$n$. In this case, as the signal concerns the current
generation and its previous one only, there cannot be more than the occurrence of two
signals belonging to different generations. In this case, the basis contains two silver
signals: that which it generates and that which it receives from the next generation.
It can be realized in the tiles by giving different horizontal channels to these
signals: an upper channel for the silver signal of the same generation, as it will go
up; a lower channel for the silver signal coming from the next generation as it will go
down. In this way, the signals of different triangles do not intersect. Consequently,
as the intersection of both directions of a silver signal of the same generation stop
each other:
    
\begin{cor}\label{uniquebeta}
In any mauve triangle of a positive generation, there is a single $\beta$-point on 
each leg.
\end{cor}

\vskip 5pt
\ifnum 1=0 {
For a reason which will appear in the next 
sections, especially about the path filling up the plane, we need to define another point
on the leg of a mauve triangle of positive generation: the {\bf $\gamma$-point}. By 
definition, the $\gamma$-point on the leg of a mauve triangle~$T$ of the generation~$n$+1
is the intersection of the $\beta$-cline defined by the hat of~$T$ with the considered leg
of~$T$. The $\gamma$-point lies in between the vertex of~$T$ and the $HP$ of~$T$.
The distance of the $\gamma$-point from the vertex is the same as the distance from 
the $LP$ and the $\beta$-point. The reason is that both are determined by $\beta$-clines
associated to triangles of the generation~$n$.

   Now, the construction of the $\gamma$-point is also rather easy.

   A signal, call it the {\bf $\gamma$-signal}, starts from the $HP$ of~$T$ and goes up 
towards the vertex of~$T$. The first basis of mauve triangle which is met by the signal
is the basis of the hat of~$T$, say $T_{-1}$. The signal goes along this basis in the 
direction of the corner of~$T_{-1}$ which is the closest to the $HP$. When the corner is 
reached, the $\gamma$-signal finds there the $\beta$-signal of the hat and it goes back,
following this~$\beta$-signal until it find the $\beta$-cline of the hat, say~$\iota$. 
It is the expected one. The $\gamma$-signal follows~$\iota$ until it meets again the leg
of~$T$: the single one which contains the $\gamma$-signal issued from the $HP$ of this
side of the triangle. Note that this signal cannot be confused with other signals belonging
to other legs. Indeed, other legs can be present but they are in a part which goes from 
their vertex to an intersection which is the mid-point of the interval going from the
vertex to the high-point, or in a part which is below their mid-point. In both cases,
there is no $\gamma$-signal in this parts of legs sot that the single leg accompanied
by a $\gamma$-signal which can be crossed by~$\iota$ is that of~$T$.

   From this we conclude that the $\gamma$-point also can be defined by a finite set of 
prototiles.
} \fi

\vskip 5pt
About the construction of the $\beta$-points, it can be 
noticed that algorithm~\ref{betapoint} can be processed 
in the reverse order. This means that it can be 
constructed from a mauve triangle~$T$ of type~2 and of 
the generation~$n$ for looking at the $\beta$-point of 
the mauve triangle~$M$ of the generation~$n$+1 which 
contains~$T$ if any. The algorithm may detect if~$M$ exists
or not and, when it exits, how to find the $\beta$-point.
The existence of~$M$ is equivalent to the presence of a 
basis at the $LP$'s of the mauve triangle of type~3 of the
generation~$n$ reached by the pre-signal emitted from~$T$
looking after such a basis. If the basis is reached, the
pre-signal returns to~$T$ in order to trigger the signals of 
algorithm~\ref{betapoint} in a reverse order which allows 
to detect the $\beta$-point. Accordingly, this provides an iterative and
bottom-up version of algorithm~\ref{betapoint} for the construction of 
the $\beta$-points.
\vskip 5pt
   A last feature about the $\beta$-point is that it allows to differentiate the part
of the $\beta$-cline of type~2 on which it lies which is contained in the triangle from
the part which is outside. 
 
   Later, we shall see that this differentiation is very important. It can easily be
realized, for instance as follows, according to the differentiation between open and covered
basis in the interwoven triangles. Each $\beta$-point emits a horizontal signal on its
isocline, outside the triangle to which it belongs. The signal is lateral and has the 
laterality of the leg. In between two consecutive mauve triangles on the same primary
latitude and of the same generation, the signals emitted by the opposite $\beta$-points
meet thanks to a join-tile which is similar to those used with the interwoven triangles.
On the part where the horizontal signal is present, we shall say that the $\beta$-cline
is {\bf covered}. In the part where it is not present, we shall say that the $\beta$-cline
is {\bf open}. Clearly, the $\beta$-cline is open inside the mauve triangles of its 
generation and it is covered in-between two consecutive such triangles within the same 
latitude.
  

\ifnum 1=0 {
   We have seen that the primary latitude of a mauve triangle is the set of isoclines
which cross its legs, the isocline of the vertex being excluded but that of the basis begin
included.

   The problem with this definition is that a triangle~3 $T_1$~of the generation~$n$ 
attached to a triangle~$T_0$ of the generation~$n$+1 is not contained in~$T_0$. However,
what is not inside~$T_0$ is the the fourth part of~$T_1$: in fact, the basis of~$T_0$
crosses~$T_1$ along the isocline of its mid-points. In some sense, we would like to
consider~$T_1$ as contained in the space defined by~$T_0$. To be coherent, this requires
to rule out the hat of~$T_0$. 

   In this way, this compels us to include in this space a part which belongs to another
primary isocline of the same generation. This is why we define the notion of {\bf latitude}
as a more complex notion as that of a primary latitude. This is also motivated by the
way for the path which is constructed in the next section.

   Let us go back to the previous mauve triangles~$T_0$ and~$T_1$, where $T_1$ is not
completely contained in~$T_0$. Now, if we look at~$T_1$ and if~$n>0$, we can repeat 
with~$T_1$ the considerations which we did with~$T_0$. There is a triangle~$T_2$ of the
generation~$n$$-$1 which has its vertex inside~$T_1$ but which is not completely contained
in~$T_1$. Moreover, the basis of~$T_1$ cuts the legs of~$T_2$ at their low point.
We can repeat this process until we arrive to a mauve triangle of the generation~$0$.
We define the latitude by the notion of {\bf border line}.

   For a given primary latitude of the generation~$n$, the {\bf border line} is a broken 
line which, in between two consecutive mauve triangles of the generation~$n$ and
within the latitude runs along the isocline of the low points of these triangles.
When the border line meets a triangle~$T_0$, it first goes down along the leg of~$T_0$
until it meets the corner. There, it goes on on the basis of~$T_0$, until it meets
the leg of a new triangle~$T_1$, necessarily of the previous generation. As the basis 
of~$T_0$ is on the isocline of the low points of the legs of~$T_1$, we can repeat
the process. When a triangle of the generation~$0$ is met, the process stops and the 
border line goes up. But it goes horizontally as soon as a basis of a triangle is met
and it can go down again when a triangle of a smaller generation is met. After
several ups and downs, the border goes back to the other lower point of~$T_0$ and, from
there, it goes on on the isocline of the low points of~$T_0$ until it meets the next 
triangle of the generation~$n$, $T'_0$. Of course, before meeting~$T'_0$, the border
line may meet other triangles of smaller generations. As it will be at their
low points, the same process as mentioned above also applies.
  
   In fact, the border line applies the following simple algorithm.

\begin{algo}\label{algoborder}
Algorithm for the border line of a given generation.
\end{algo}
\vspace{-12pt} 
\grostrait
\ligne{\hfill
\vtop{\leftskip 0pt\parindent 0pt\hsize=250pt
{\bf loop}\\
\ligne{\hskip 24pt {\bf if} $isocline$\hfill}
\ligne{\hskip 36pt {\bf then} $go on$;\hfill}
\ligne{\hskip 36pt {\bf elsif} $outer leg$;\hfill}
\ligne{\hskip 48pt {\bf then} $go down$;\hfill}
}
\hfill}
\demitrait

} \fi

\subsection{The latitude}

\ifnum 1=0 {
It is interesting to know
more about the intersections, in general terms, between mauve triangles.

   However, lemma~\ref{innerandLP} gives us a very important information. Consider a mauve
triangle~$T$ of the generation~$n$+1 and consider four triangles~$T_i$ of the 
generation~$n$, $i\in\{0..3\}$, which are inside~$T$, where $T_0$ is within the latitude 
which is just below the hat of~$T$, and~$T_{i+1}$ is just below the latitude of~$T_i$
for $i\in\{0..2\}$. We generalize these notions to all mauve triangles of a given
generation compared with the next one. We shall say that a triangle~$M_0$ of the 
generation~$n$ is an $i$-triangle in a triangle~$M_1$ of the generation~$n$+1,
with $i\in\{0..3\}$ if the latitude of~$M_0$ is in the same position with respect 
to~$M_1$ as the latitude of~$T_i$ with respect to~$T$. Note that the isocline of the 
low points of~$T$ is that of the low points of~$T_2$ and that the basis of~$T$ is on 
the isocline of the low points of~$T_3$.
} \fi

   From lemma~\ref{innerandLP}, we know the intersections between mauve triangles of
the generation~$n$+1 and those of the generation~$n$. We have to look at a more general
situation.

   From the construction of the interwoven triangles, we know that the bases and vertices
of mauve triangles characterize the corresponding triangles. This is not the case for the
isocline of their~$LP$'s: such an isocline is the mid-distance line of phantoms.
We know that usually, the mid-distance line of a phantom may be such a line for several
phantoms of different generations. It is the reason why the same ambiguity is attached to 
the isoclines of the $LP$'s as we can see from lemma~\ref{innerandLP}. Recursively applying 
the lemma to inner triangles in a fixed mauve triangle, we obtain that $LP$'s of a 
triangle of the generation~$n$ may be crossed by the basis of a triangle of the 
generation~$n$+$k$, for any positive~$k$. In general, it is not possible to predict 
if such a situation will occur. Now, if it occurs, we know that inside the 
mauve triangle of the generation~$n$, the 2-triangles will also be cut by this basis, 
also at their $LP$'s.

   From lemma~\ref{innerandLP}, we know that this situation does not occur for the
0- and 1-triangles which are contained in a mauve triangle. These triangles may be
intersected by smaller triangles only, which cut their basis or their legs near their
vertices.

    Going back to 2-triangles, we can see that if a 2-triangle~$T$ is of a generation~$n$ 
with~$n>0$, we can find smaller triangles which are also 2-triangles inside~$T$,
their legs being cut by the basis of~$T$, at their $LP$'s too. And this can be repeated 
until we reach the generation~0.

    We can say the same for~3-triangles. If such a triangle is not of the 
generation~0, its basis cuts triangles of the previous generation, and this property can 
be repeated recursively. Remember the notion of shadow of a triangle and the construction
of the $\beta$-cline.

\ifnum 1=0 {
\vskip 7pt
   From what we have seen, remark that a 2-triangle may be cut and may be not. As an
example, a 2-triangle of the generation~$n$ which is inside a 0- or a 1-triangle
of the generation~$n$+1 is never cut. However, even if it is in a 3-triangle of the 
generation~$n$+$k$, with $k\geq 1$, it is always cut: as the triangle~$K$ of the 
generation~$n$+$k$ which we consider exists, all the inside triangles also exist, in 
particular the 3-triangles of the generation~$n$+$k$$-$1. Now, the basis of~$K$ cuts 
the basis of the 3-triangle of the generation~$n$+$k$$-$1 and, recursively, all the 
2-triangles which are inside one another, starting from the 2-triangles of that 3-triangle. 
} \fi

   From this, we define the {\bf border line} of a primary latitude of the generation~$n$ 
as a broken line as follows:

   First, define the {\bf bottom} of a mauve triangle as the broken line which consists
of the legs of the triangle from the $LP$ to the corner and the basis.
  
Then, we define the {\bf border line} as the isocline of the $LP$'s of the triangles
of the generation~$n$ of this primary latitude in which each maximal segment which falls
inside a mauve triangle~$M$ of a generation at most~$n$$-$1 is replaced by the bottom 
of~$M$, the same process of substitution being recursively applied to the basis of the 
triangle and of the substituted triangles. The term maximal indicates that we take the 
biggest triangle of a generation at most~$n$$-$1 which is cut by the isocline.

   From now on, the {\bf latitude} of a triangle of the generation~$n$ is the set of
tiles which is contained between the border line of its primary latitude and the border
line of the same generation which is attached to the primary latitude which is just above.
We include all the tiles of the lower border and we include none of the upper border.

   Note that in a border line, when we apply the recursive process of substitution of
bottoms of triangles starting from a triangle of the generation~$n$, the bases which are
the further from the isocline of the $LP$'s are bases of the generation~$0$. They are
all on the same isocline which we call the $\beta$-{\bf cline}.

   Now that the notion of latitude is clearly defined, let us look at
what happens between two consecutive triangles~$T_1$ and~$T_2$ of the same generation 
which belong to the same latitude.

   A priori, we have three situations:

{\leftskip 20pt\parindent 0pt
$(i)$ for both~$T_1$ and~$T_2$, the vertex does not belong to a basis of a mauve triangle;

$(ii)$ the vertex of~$T_1$ does not belong to the basis of a mauve triangle but the
vertex of~$T_2$ does;

$(iii)$ each vertex belongs to a basis of a mauve triangle.
\par}   

   In fact, we have:

\begin{lem}\label{samevertices}
Consider two mauve triangles~$T_1$ and~$T_2$ of the generation~$n$ and belonging to the
same latitude. Assume that $T_1$~and $T_2$~are consecutive. Then, if the vertex of~$T_i$
belong to the basis of a triangle~$B_i$ for $i\in\{1,2\}$, then $B_1=B_2$.
\end{lem}

\vskip 5pt
\noindent
Proof. 
Assume that $T_1$ and $T_2$ have both their vertices on a basis of 
triangles~$B_1$ and~$B_2$ of the same generation as stated in the lemma. Assume
that $B_1$ and~$B_2$ are distinct. From what we have noted in section~2.1, as $T_i$ 
has its vertex on the basis of a mauve triangle~$B_i$, $i\in\{1,2\}$,
then $B_i$ contains the remotest ancestor~$A_i$ of~$T_i$. We also know that the vertex 
of~$A_i$ is at a distance at most the fourth of the height of~$T_i$.
From this, the vertices of~$A_1$ and~$A_2$ are below the mid-distance line of~$B_1$ 
and~$B_2$. And so, if we assume that $B_1\not= B_2$, even if 
the vertices of~$B_1$ and~$B_2$ are close to each other, the vertices of~$A_1$ and~$A_2$ 
are so far from each other that on the isocline which joins them there is another 
vertex of a mauve triangle~$A_3$ in between them. Now, as all mauve 
triangles generated from~$A_3$ exist, there is a mauve triangle~$T_3$ of the same 
generation as~$T_1$ which stands in between~$T_1$ and~$T_2$. A
contradiction. And so, $B_1=B_2$. \cqfd

\subsection{The $\gamma$-points and the high points}

   We conclude this section with the notion of {\bf $\gamma$-point} and of
{\bf high point}, $HP$ for short, which both play an important role in the next section.

   Intuitively, the $LP$~corresponds to the entry of the path into a triangle and the~$HP$
corresponds to its exit. The $\gamma$-point plays a similar role to that of the 
$\beta$-point.

\subsubsection{The $\gamma$-point and its construction}

   The $\gamma$-point is defined by the intersection of the leg of mauve triangle
with the $\beta$-cline defined by its hat, if any. The difficulty comes from the
fact that the hat may not exist while the $\gamma$-point can always defined for a
mauve triangle of a positive generation.

   As for the $\beta$-point, the $\gamma$-point is not defined for a mauve-0 triangle.
For a mauve triangle~$T_1$ of generation~1, consider the above definition when the
hat exists. We remark that the $\beta$-cline is defined by the basis of the hat as it is
a mauve-0 triangle. Now, the basis of the hat contains vertices of the 0-triangles
of generation~0 contained in~$T_1$. Now, as $T_1$ exists, its inner 0-triangles also
exist. And so, it is possible to define the $\gamma$-points of~$T_1$ by using its
0-triangles only.

   First, we call {\bf first points}, $FP$ for short, the point of a leg of a mauve 
triangle~$T$ which is at a distance~$\displaystyle{h\over4}$ from the vertex of~$T$.
The construction of the $\gamma$-points starts by the determination of the $FP$'s which is
easy: it is not difficult to see that the $FP$'s of~$T$ are the mid-points of the red
triangle~$R$ whose vertex is the same as that of~$T$.

   Then, we proceed as follows:
\vskip 5pt
\grostrait

\vskip 2pt
{\leftskip 40pt\parindent-20pt\rightskip 20pt
two $\gamma$-signals start from the $FP$ of~$T_1$: one to its vertex, along the leg,
the other inside the triangle;

the inside signal goes on along the isocline until it meets the closest 0-triangle~$M_0$
to this leg of~$T_1$; there, it goes up along the leg until it reaches the vertex
of~$M_0$; the $\gamma$-signal goes back to the leg of~$T_1$, following the isocline
of the vertex of~$M_0$;

the intersection of the $\gamma$-signal going back to the leg with the $\gamma$-signal
going up along the leg defined the $\gamma$-point of~$T_1$.
\par}

\demitrait
\vskip 5pt
   The justification of this algorithm relies on the fact that the isocline~$\iota$ which
joins the $FP$'s of~$T_1$ passes through the $LP$'s of the 0-triangles which are
contained in~$T_1$. From this fact which is easy to establish, and from the intersection
properties of the mauve triangles, we conclude that all other mauve triangles which
are encountered by the isocline~$\iota$ are 2-triangles, and that $\iota$ passes by the
$LP$'s of these triangles too.

   The general case is not much more difficult to establish by the following
recursive algorithm.

\vtop{
\begin{algo}\label{gamma}
\leurre
The construction of the $\gamma$-point of a triangle~$T$ of the generation~$n$$+$$1$.
\end{algo}
\vspace{-5pt}
\grostrait
\vskip 2pt
{\leftskip 40pt\parindent-20pt\rightskip 20pt
two $\gamma$-signals start from the $FP$ of~$T$, one along the leg towards the vertex
and the second inside the triangle along the isocline which joins the $FP$'s;

the inside signal goes on until it meets the first 0-triangle~$M_0$ inside~$T$; there,
meeting~$M_0$ at an $LP$, it goes up along the leg of~$M_0$ until it reaches the 
$\gamma$-point~$G_0$ of~$M_0$; there, it goes back to the leg of~$T$, on the
isocline which passes through~$G_0$, circumventing the inner triangles which
it encounters;

the intersection of the $\gamma$-signal going back from~$G_0$ to the leg of~$T$
with the $\gamma$-signal climbing along this leg defines the $\gamma$-point of~$T$;
the $\gamma$-point stops both $\gamma$-signals.  
\par}

\demitrait
}
\vskip 5pt
   The justification of the construction given by algorithm~\ref{gamma} is provided by
the following lemma.

\begin{lem}\label{FPline}
   Let $T$~be a mauve triangle of the generation~$n$$+$$1$. The isocline which passes
through its $FP$'s encounters mauve triangles inside~$T$ of types~$0$ and~$2$ only.
The meeting occurs at the $LP$'s of the inner triangles. The encountered $0$-triangles 
belong to the generation~$n$. The encountered $2$-triangles belong to a generation~$i$
with $i<n$.
\end{lem}

\noindent
Proof. The conclusion on the generations of the inner 0- and 2-triangles encountered by
the isocline~$\iota$ which passes through the $FP$'s of~$T$ is a consequence of the fact that
the meeting with the legs of the triangles occur at their $LP$'s, as we have seen 
in lemma~\ref{innerandLP}. Now, the fact that $\iota$~meets the inner triangles at 
their $LP$'s is easy. The 3-triangles of the generation~$n$ inside~$T$ are cut by the 
basis of~$T$ at their~$LP$'s. Now, the distance in isoclines from the basis 
of~$T$ to its~$FP$'s is $\displaystyle{{3h}\over4}$ where 
$h$~is the height of~$T$. Now, the height of a triangle of the generation~$n$
is~$\displaystyle{h\over4}$. Accordingly, for what is the disposition of the isoclines,
$\iota$~is in the same position with respect to a 0-triangle of the generation~$n$ as 
the basis with respect to a 3-triangle of the generation~$n$. Consequently, inside the 
0-triangle of the generation~$n$, the isocline meets inner triangles at their $LP$'s 
and they are all of the type~2 as already seen. By synchronization,
it is the same for the inner triangles, encountered by this isocline. \cqfd

   We have an additional interesting property:

\begin{lem}\label{gammaaligned}
The isocline of the $\gamma$-points of a mauve triangle meets other mauve triangles
at their $\gamma$-points too.
\end{lem}

\noindent
Proof. This comes from the structure of the tower which defines a $\beta$-cline.
And, by construction, the isoclines through the $\gamma$-points is a $\beta$-cline.
\cqfd
\vskip 5pt
We can formulate the same remark about 
algorithm~\ref{gamma} as the one which was formulated for 
algorithm~\ref{betapoint}. The construction can also 
performed in the reverse order. Again we have a pre-signal
which detects the existence of a containing mauve triangle
of the next generation. It is the same signal as 
previously, looking after a basis at the $LP$ of a mauve
triangle of type~3 reached from the considered mauve 
triangle of type~0. If the basis is found, the pre-signal
goes back to its emitting point in order to trigger the
signals of algorithm~\ref{gamma} in the reverse order.
Again, this provides us with an iterative and bottom-up
version of algorithm~\ref{gamma}.
\vskip 5pt
\subsubsection{The $HP$}

   From the notion of $\gamma$-points, it is easy to define the $HP$'s.

   Indeed, the $HP$'s of a mauve triangle~$T$ is defined by the following construction.

   A signal starts from each $FP$ of~$T$ and goes up along the leg, towards the vertex
of~$T$. If there is a $\gamma$-point, then if the $\beta$-cline which passes through
the $\gamma$-point is a $\beta$-cline of type~2, the $HP$ is the $\gamma$-point
and the signal stops there. Otherwise, the signal goes on climbing along the vertex
until it meets the first basis which cuts the legs of~$T$ if any. If such a basis is
encountered, the meeting with the legs of~$T$ define the $HP$'s. If not, the $HP$ is
the tile of the leg which is on the isocline which is just below the vertex. This is
also the definition of the $HP$ for a mauve-0 triangle.

\section{An almost plane-filling path}

   Now, we turn to the construction of the path. The general strategy which we follow
was presented in~\cite{mmCSJMprefill,mmarXivprefill}, but we shall make it much more 
precise.

   The path goes from an~$LP$ to a~$HP$ and then to an~$LP$ and so on. It can be seen 
as a bi-infinite word of the form $^\infty((LP)(HP))^\infty$ on the alphabet $\{LP,HP\}$.

   Roughly speaking, we fill up a latitude until we meet legs which cross both the upper 
and the lower border of the latitude. Then, we go up or down, depending on the direction of
the path and into which type of basic region we fall: the type of a bigger triangle 
or of a zone in between two bigger triangles.

   In most cases, this strategy is enough to fill up the whole plane.

   Later, we shall discuss about the exceptional cases.
 
\subsection{The regions and the path}


   Our first task is to define the regions which we shall investigate and then, how the
path is built on the basis of what will be called the {\bf basic regions}.

   We have two basic regions. The first one is the set of tiles defined by a mauve 
triangle: its borders and its inside. Remember that the basis of a mauve triangle 
contains more than the majority of tiles resulting from the just given definition. 
It is considered as a basic region as once the path enters a mauve triangle~$T$, it 
fills up~$T$ almost completely before leaving~$T$. In fact, there is a restriction and
the path fills a bigger area. In fact, the path also fills up the space which
is contained between the basis of~$T$ and the part of the border of the latitude of~$T$
which is delimited by the corners of~$T$, the tiles on this border being included. The 
restriction comes from the definition of the latitude: we have to withdraw at least
the tiles belonging to the border of the just upper latitude of the same generation.
An additional restriction comes in the case when the $HP$ is on an open $\beta$-cline 
of type~2, as we shall describe this later.

   The other type of a basic region is defined by the area in between two consecutive
mauve triangles of the same generation within the same latitude.

    We already know that the just indicated regions
can be split into four horizontal slices defined by the types of the triangles of the just
previous generation which are contained in these regions. Now, if we go from one side to
another in each slice, and if the directions alternate from one slice to the next one,
this even number raises a problem: a priori, starting from one side, we go back to the
same side. To solve this problem, we split one slice into two ones thanks to the
$\beta$-cline of type~2: inside a mauve triangle, there is a unique open $\beta$-cline 
of type~2 which runs from one leg of the triangle to the other. It is the isocline of the
$\beta$-points. This $\beta$-cline splits the region of type~3 into to sub-slices. 
We shall use the second one to go back to the original side. As there remain three 
slices, we go from the original one to the opposite one, as required.

   This is the general principle for defining the path. Note that this principle
holds both for triangles an the in between region. We shall now turn to the precise
description.

   We shall examine how we fill up the basic regions for generation~0 and we shall then 
proceed by induction from~$n$ to~$n$+1. In fact, as we shall see, the induction step
can be based on what is to do for the basic regions of generation~1.

\vskip 7pt
\noindent
$\underline{\hbox{For generation~0}}$      
\vskip 7pt
   The situation of generation~0 is the simplest one.

   For a triangle, the trajectory of the path is the following. The path enters the
triangle through one of its $LP$'s, say~$A$. Then, it runs along the leg of the triangle,
downwards, until it reaches the corner. On this way, the path is in the inside part of the
tile which supports the leg. At the corner, the path follows the basis, until it reaches the
other corner. There, it goes up along the leg to the next isocline and there, it goes
along the isocline to the leg of~$A$. Just before reaching the leg, the path goes up to 
the next isocline and there, it runs along it until it reaches the leg, opposite to~$A$.
This back and forth motion, climbing up by one isocline each time a leg is reached goes on
until the path reaches the top of the triangle. There, the path exits from the triangle
through the isocline~$-$1 below the vertex or the isocline~$-$2, depending on the type
of the triangle: if the triangle is of type~3, the path exists through the isocline~$-$2,
if not, it exits through the isocline~$-$1. The exit~$B$ is placed on this isocline, on 
the leg of the triangle which is opposite to the leg on which $A$~lies. 
Figure~\ref{mauve_0} illustrates
this part of the path for a triangle when the topmost isocline is not occupied by another
segment of the path.

\vskip-20pt
\setbox110=\hbox{\epsfig{file=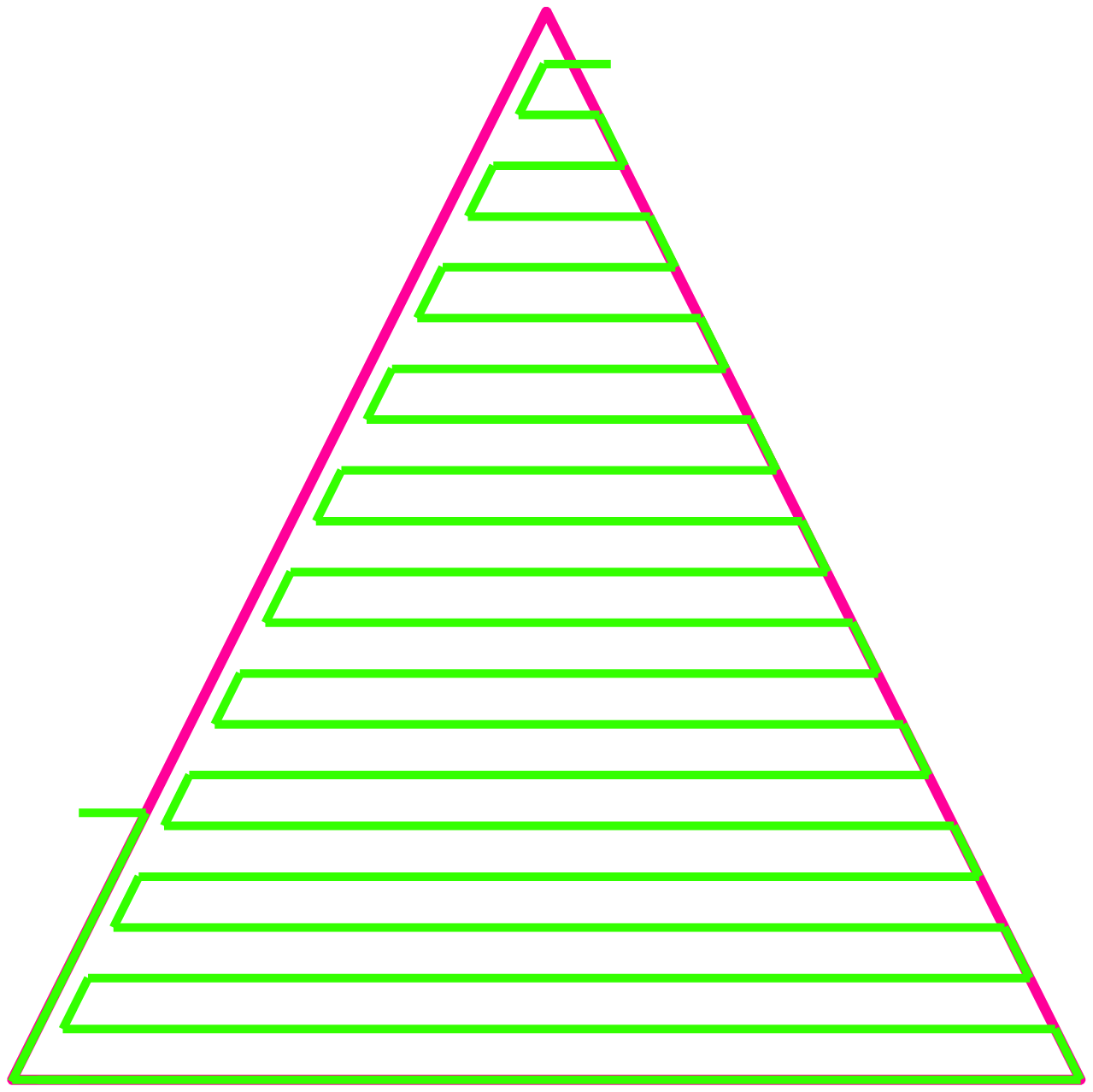,width=220pt}}
\vtop{
\ligne{\hfill
\PlacerEn {-120pt} {0pt} \box110
\hfill}
\vspace{-25pt}
\begin{fig}\label{mauve_0}
\leurre
The path inside a triangle of generation~$0$.
\end{fig}
}
\vskip 7pt
   For a part between two consecutive mauve-0 triangles within the same latitude,
we have the three situations which result from lemma~\ref{samevertices}.

   The easiest situation is when two consecutive mauve-0 triangles have their vertices
on the basis of the same mauve-0 triangle. In this case, we have a similar zig-zag line
as in a triangle. Figure~\ref{trapeze_0} illustrates this situation which looks like
very much to what happens in a mauve-0 triangle.
\vskip-10pt
\setbox110=\hbox{\epsfig{file=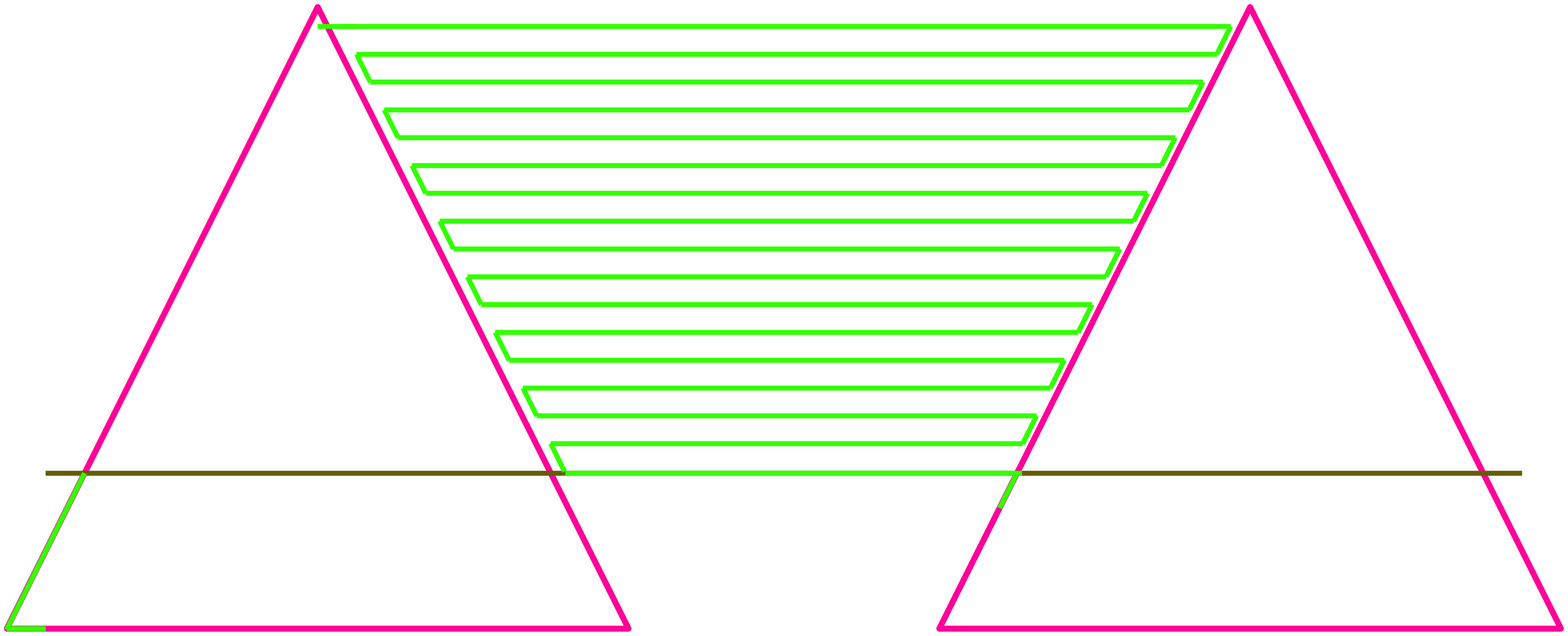,width=280pt}}
\vtop{
\ligne{\hfill
\PlacerEn {-140pt} {0pt} \box110
\hfill}
\vspace{-5pt}
\begin{fig}\label{trapeze_0}
\leurre
The path in between two consecutive triangles of generation~$0$ within the same
isocline.
\end{fig}
}
\vskip 7pt
\vskip-10pt
\setbox110=\hbox{\epsfig{file=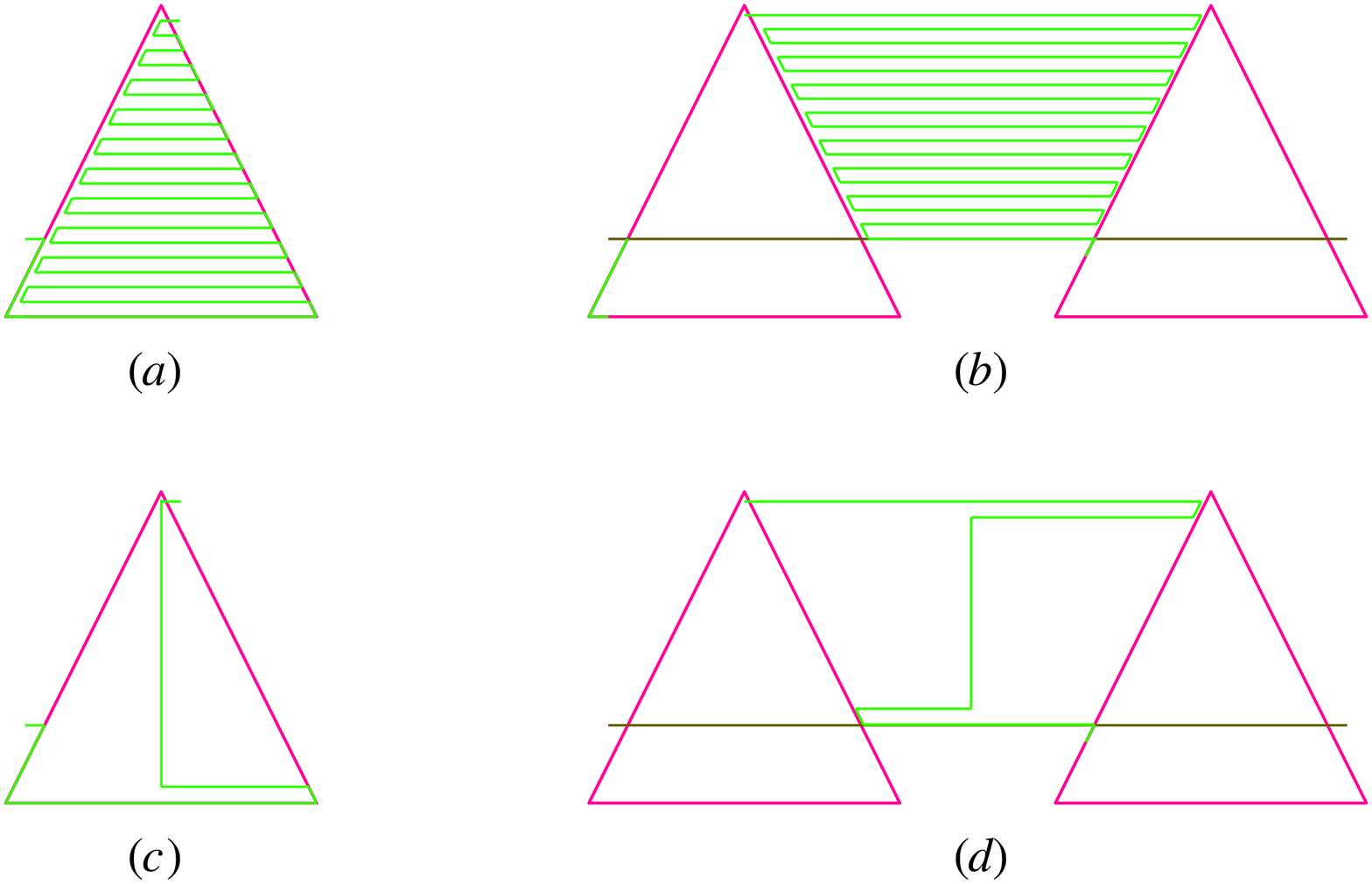,width=280pt}}
\vtop{
\ligne{\hfill
\PlacerEn {-140pt} {0pt} \box110
\hfill}
\vspace{-5pt}
\begin{fig}\label{schemas}
\leurre
A schematic representation of the path:
\vskip 0pt
\noindent
On the left-hand side, inside a mauve-0 triangle. On the right-hand side, 
in between two consecutive mauve-0 triangles within the same latitude when the
vertices belong to the same basis.
\end{fig}
}
\vskip 7pt

   In order to describe what happens in the other situations, we define a schematic 
representation of the zig-zag path of figure~\ref{mauve_0} and~\ref{trapeze_0}
in figure~\ref{schemas}. Now, as these situation will be involved starting from
generations with a positive number, we postpone the representation of the other
case of basic regions of generation~0 to the situation 
concerning generation~1.

   It is important to indicate that the representations of figure~\ref{mauve_0} 
and~\ref{trapeze_0} are also schematic. In fact, the actual trajectory of the path
is a bit more complex. We cannot decide that on a leg of a mauve triangle of any generation
the path strictly goes on the tiles crossed by the mauve signal and only them. 
If we do this, we cannot have a path which goes through any tile according to the indicated
scenario. However, it is possible to slightly change the trajectory of the
path in order to make things possible. If the path has to go along the leg, there is no 
problem to organize the hairpin-turns of the zig-zags when the path is on the left-hand 
side of the leg. Indeed, the leg is along a line of black nodes. There is a 'parallel'
line of white nodes which are the left-hand side neighbours of the black nodes. 
When the path is on the left-hand side, the organization of the hairpin-turns requires 
some care, otherwise, the condition to go through all tiles once would not be respected.
The path on the leg follows the following pattern. Let $\alpha_k$ denote the black nodes
of the leg, $k$ indicating an absolute number for the isocline on which the node is.
Let $\beta_k$ be the right-hand side neighbour of~$\alpha_k$ and let~$\gamma_k$
be the right-hand side neighbour of~$\beta_k$. Note that $\beta_k$ is the white son 
of~$\alpha_{k-1}$ and that $\gamma_k$ is the black son of~$\beta_{k-1}$. From
the connections between nodes of a Fibonacci in the heptagrid, we know that there is no
common edge between~$\beta_{k-1}$ and~$\beta_k$. A solution is the following path:
go from $\alpha_k$ to $\alpha_{k-1}$, then to~$\beta_{k-1}$ and then to~$\alpha_{k-2}$ 
and from there, repeat the just described pattern. The pattern of hairpin-turns for 
an inside path is defined as follows, using the same notations. The path arrives at 
$\beta_k$ from~$\gamma_{k+1}$. It goes to the right from~$\beta_k$ on the isocline~$k$.
When it arrives to the leg on the isocline~$k$$-$1, it stops at~$\gamma_{k-1}$ from
where it goes to~$\beta_{k-2}$, from where it goes back to the right on the 
isocline~$k$$-$2. See the illustration provided by figure~\ref{hairpins}.

\vskip-50pt
\setbox110=\hbox{\epsfig{file=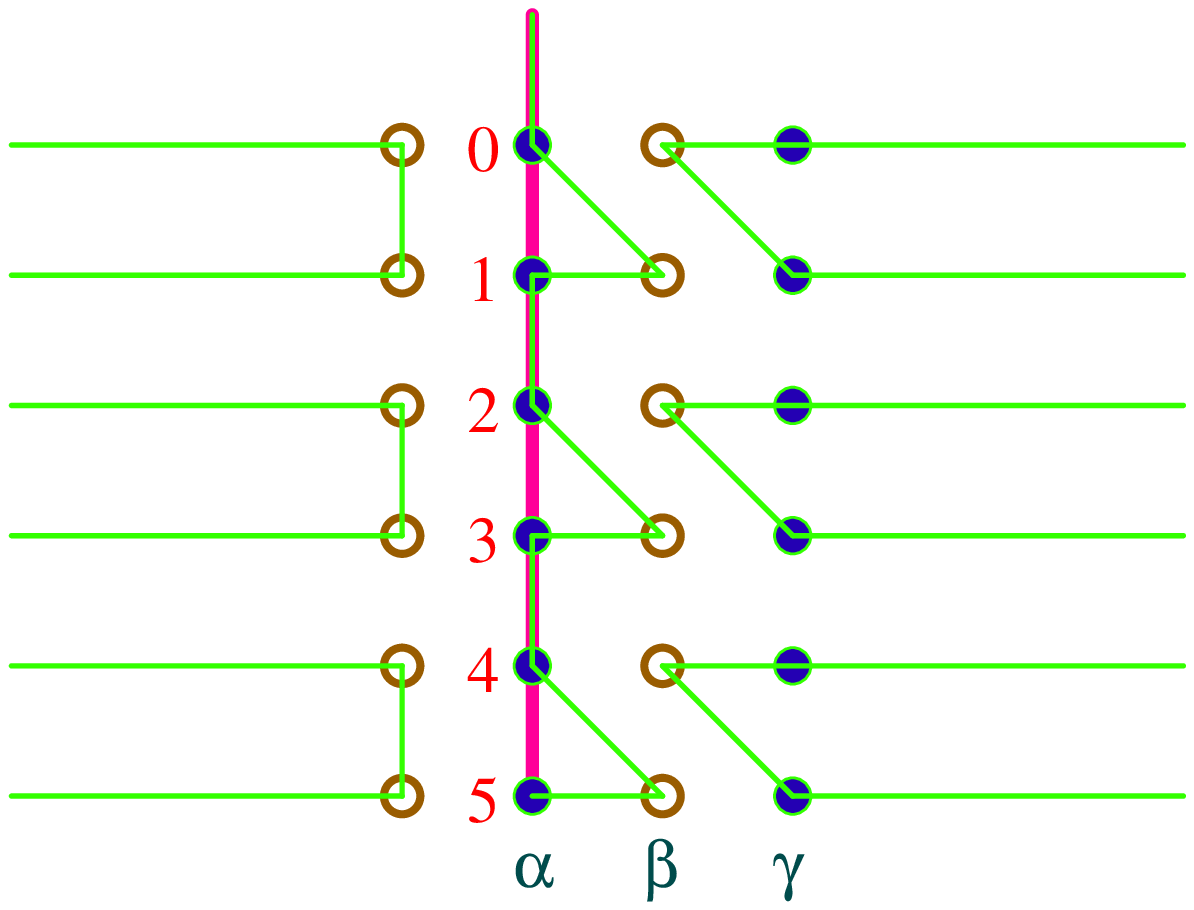,width=320pt}}
\vtop{
\ligne{\hfill
\PlacerEn {-160pt} {0pt} \box110
\hfill}
\vspace{-80pt}
\begin{fig}\label{hairpins}
\leurre
The adaption of the path close to a leg of a mauve triangle.
\end{fig}
}
\vskip-45pt
\noindent
$\underline{\hbox{For generation~1}}$      
\vskip 7pt
   First, we look at what happens in between two consecutive mauve triangles of generation~1
within the same latitude. Denote them by~$T_1$ and~$T_2$, with $T_1$~on the left-hand side
of~$T_2$. We shall assume that the path enters a mauve triangle of
generation~1 through a low point and that it exits the same triangle through its top,
on the leg which is opposite to the entry point.

   In figure~\ref{mauve_1}, we consider the case when a mauve triangle~$T_1$ of 
generation~1 is hatted by a mauve triangle of generation~0. 
The next mauve triangle of generation~1 to the right, say $T_2$ is not hatted as it 
can be easily concluded from the distance between the corners of two consecutive 
mauve-0 triangles. 
 
   There are necessarily 0-triangles of generation~0{} in between~$T_1$ and~$T_2$.
Figure~\ref{mauve_1} illustrates a schematic situation of the $i$-triangles of 
generation~0 which we may find in in between~$T_1$ and~$T_2$. Note that
we have 0-, 1- and 2-triangles. The 3-triangles are not represented as they do not
belong to the latitude of generation~1 defined by~$T_1$ and~$T_2$.

   The figure illustrates the way of the path, assuming that it exits from~$T_1$ through
its right-hand side $HP$ in order to enter~$T_2$ through its left-hand side~$LP$.
We have to take into account the behaviour of the path in the primary latitude
of~$H_1$. It again appears in the figure by looking at the configuration of the
0- and 1-triangles of generation~0 which are in between~$T_1$ and~$T_2$.

\vskip-100pt
\setbox110=\hbox{\epsfig{file=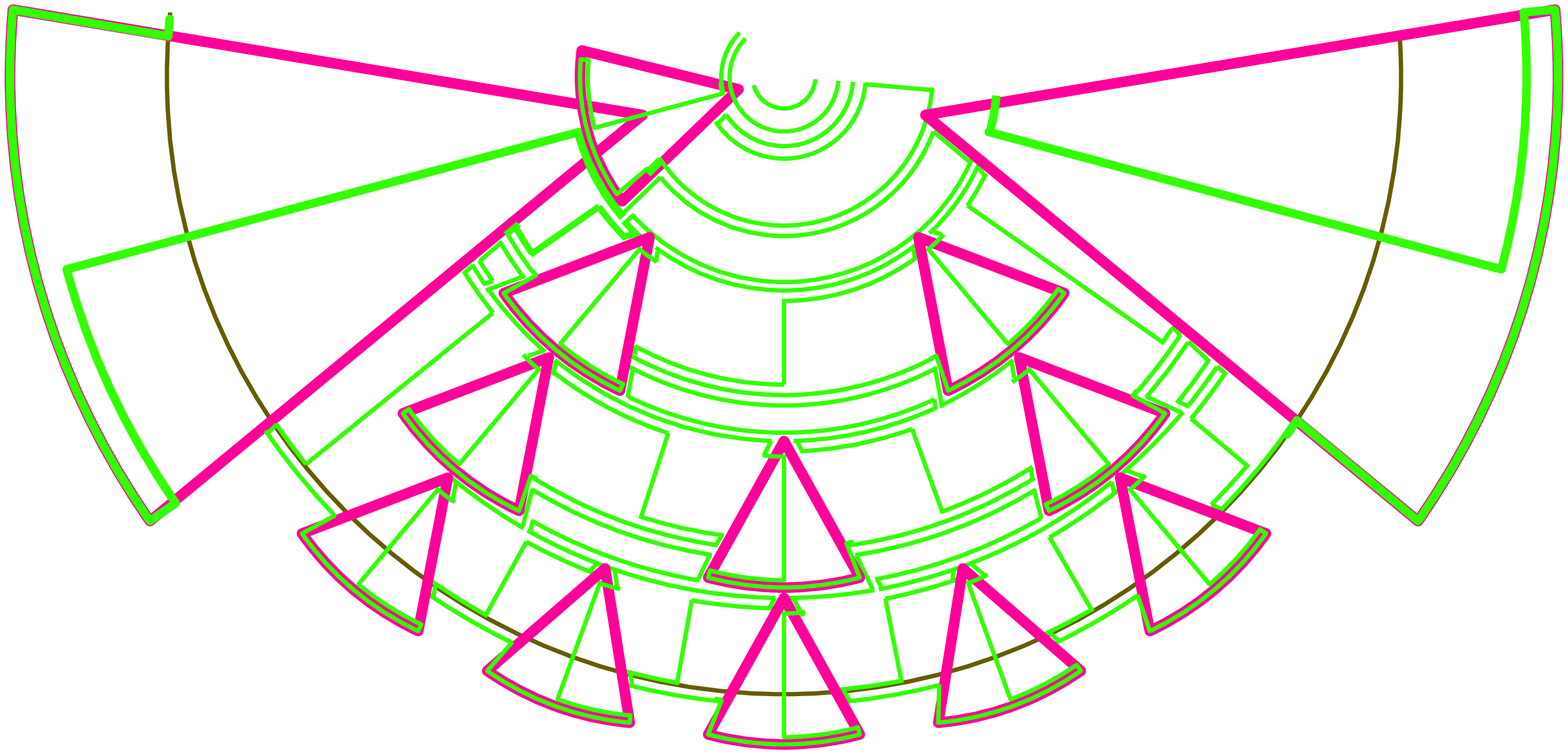,width=360pt}}
\vtop{
\ligne{\hfill
\PlacerEn {-160pt} {0pt} \box110
\hfill}
\vspace{-25pt}
\begin{fig}\label{mauve_1}
\leurre
The path in between two triangles of generation~$1$.
\end{fig}
}

\vskip 5pt
   First, the path follows the border of~$H_1$ and then climbs along its right-hand leg 
until it reaches the isocline which is just below the $LP$ of~$H_1$. It goes on
along this isocline until it reaches the left-hand side leg of~$T_2$, just below the 
vertex of~$T_2$. Next, it follows a zig-zag way until it goes back to the point~$M$
defined by the corner of~$H_1$. This point~$M$ lies on the isocline~$\iota$ which is 
just below the basis of~$H_1$ and is on the way upwards taken by the path. 
During the zig-zag, the path meets the vertices of 0-triangles of generation~0.
As the path inside a triangle never passes through its vertex, 
the path may cross them, as if it would do if a basis would contain these vertices. 
Coming back after leaving the
closest vertex of such a 0-triangle to~$H_1$ and traveling on the isocline~$\iota$+1, 
the path arrives to the tile which is before the tile of the path above~$M$ on~$\iota$+1. 
There, the path goes down to~$\iota$ and, on the tile which is adjacent to~$M$, it 
goes on the isocline~$\iota$ in the direction of~$T_2$. Now, the path does not meet~$T_2$ 
but a 0-triangle~$T_0$ of generation~0, which it reaches just below the vertex. 
Accordingly, the path zig-zags downwards, oscillating between~$T_1$ and~$T_0$. By this 
oscillating motion, the path reaches the $LP$ of~$T_0$: it enters the triangle
which it fills according to the motion defined by figure~\ref{mauve_0}. When the path 
exits from this triangle, it follows 
the way defined by figure~\ref{trapeze_0}
until it reaches the next 0-triangle on its way to~$T_2$. Accordingly, this sequence
is repeated until the last 0-triangle of generation~0 before~$T_2$. Now, when the
path exits from the triangle, it is barred by the former passage of the path on~$\beta$
and so the path goes on~$\iota$ until it reaches~$T_2$. But, as the path exited from~$T_1$
and as it is close to~$\beta$, it knows that it cannot enter~$T_2$. And so, it goes downwards
and zig-zagging. Now, during this zig-zag, it will meet the $LP$ of the last 0-triangle
of generation~0: this $LP$ is closed as the path filled up this triangle. We shall
later see the mechanism which forces one $LP$ to be open and the other to be closed.
And so, going down, still zig-zagging, the path will meet the $LP$ of the
closest 1-triangle of generation~0 to~$T_2$. Here, the $LP$ is free, so that the
path enters the triangle. 

   Now, we turn to the route of the path inside a mauve triangle of generation~1.

\ligne{\hfill}
\vskip-95pt
\setbox110=\hbox{\epsfig{file=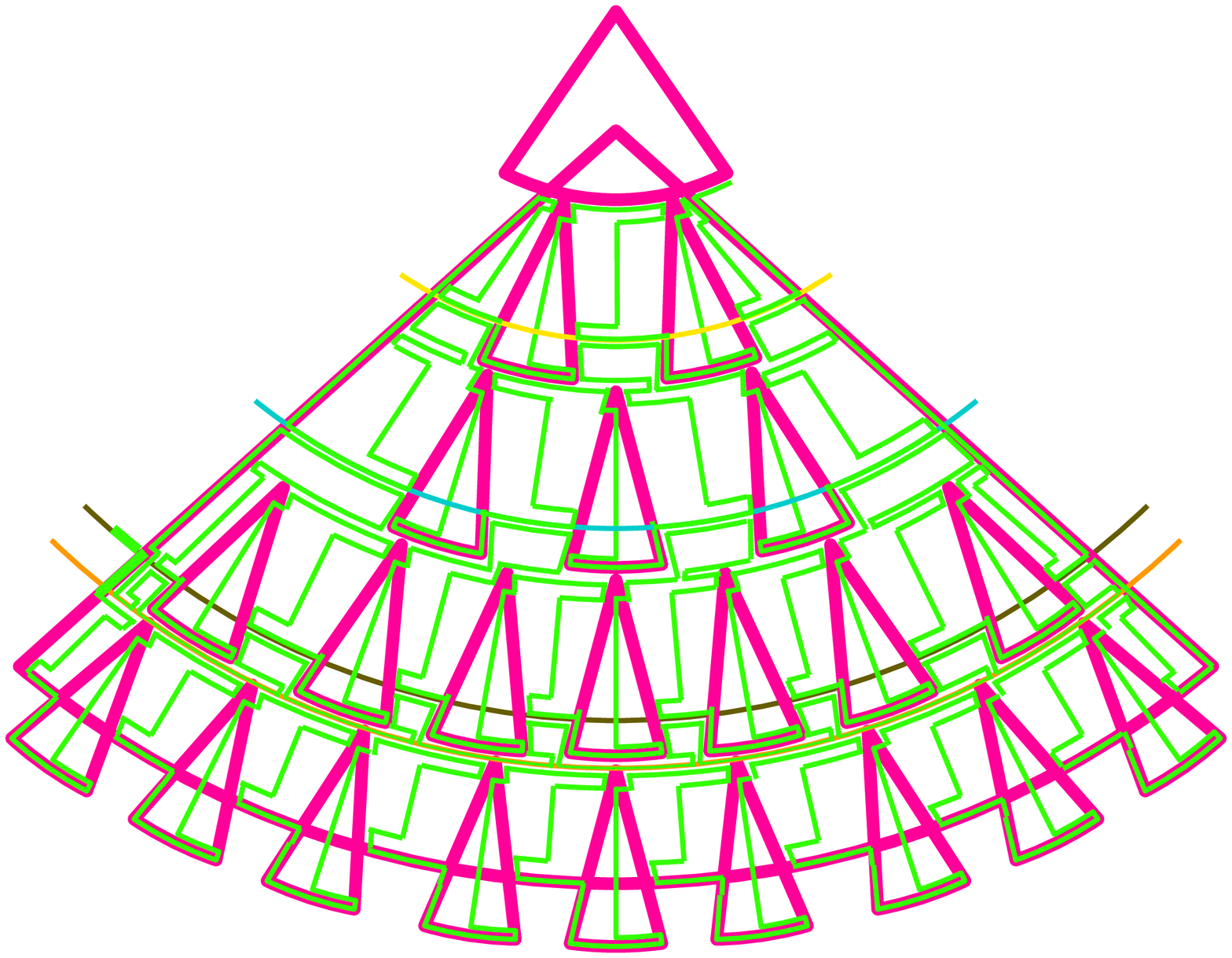,width=280pt}}
\setbox112=\hbox{\epsfig{file=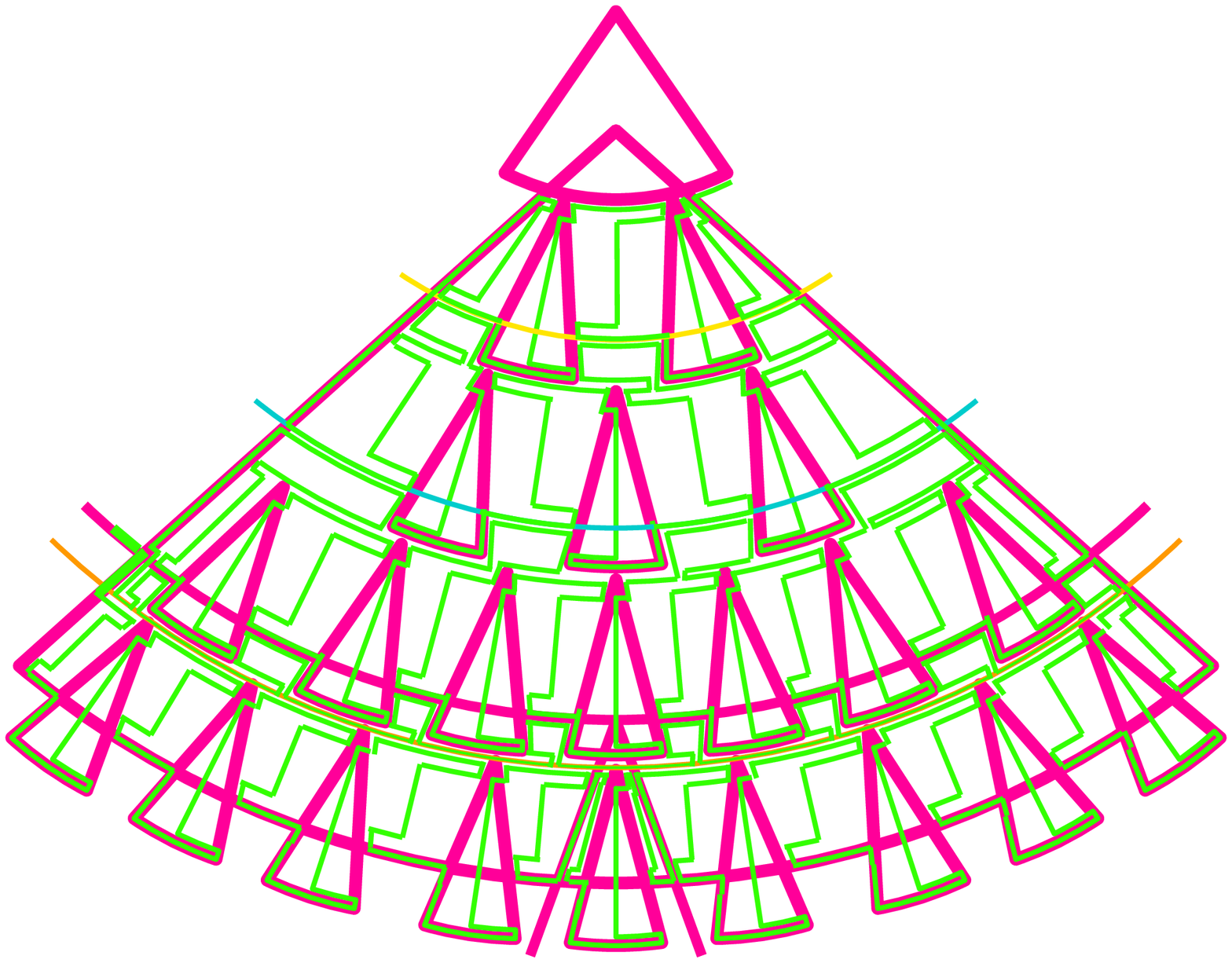,width=280pt}}
\vtop{
\ligne{\hfill
\PlacerEn {-210pt} {0pt} \box110
\PlacerEn {-40pt} {0pt} \box112
\hfill}
\begin{fig}\label{mauve_1in}
\leurre
The path inside a triangle of generation~$1$.
\vskip 0pt
On the left-hand side: a $0$- or a $1$-triangle. On the right-hand side, a~$2$- or a 
$3$-triangle which is cut by a mauve triangle of a bigger generation.
\end{fig}
}

   In both pictures of figure~\ref{mauve_1in}, we can see an open $\beta$-cline of
type~2 which cuts the strip delimited by the line of the $LP$'s and the basis of the
triangle into two parts. 

   This is a general feature. This cut allows to make the path going back near the~$LP$
through which it entered the triangle in order to cross the latitude of the 2-triangles
in the direction from~$LP$ to~$HP$, where $LP$ refers to the side of the triangle through
which the path entered and~$HP$ refers to the other side as the path will exit through
the $HP$ of this other side. The crossing of this latitude inside the triangle obeys the
same principles as in between two triangles. When arriving almost to the closed~$LP$,
the path goes up to the isocline which is below the~$FP$. From this point, it crosses
the latitude of the 1-triangles, this time in the direction from~$HP$ to~$LP$. When it
arrives to the other side, the path goes up along the leg until it arrives by one 
isocline below the $HP$ of this leg. From there, it crosses the latitude of the 0-triangles,
in the direction from~$LP$ to~$HP$. When the crossing completes, the path arrives at 
the~$FP$ from where it goes to the right~$HP$ by going up along the leg of the triangle.

   In between two consecutive mauve triangles of generation~1 within the same latitude,
the $\beta$-cline~2 which we noticed inside a triangle~$T$ of generation~1 plays a similar
role but for another latitude: for the one which is below the latitude of~$T$. Now, for a
basic regions, there are a lot of $\beta$-clines of type~2 which cross the legs of the 
triangles defining these regions. The $\beta$- and $\gamma$-points tell us which one are
important for the region: only those which pass through this points. The other intersections
are not important.

   In a basic region, we have four sub-latitudes, corresponding to the four types of 
mauve triangles of the previous generation. In order to go in the right direction, we need
to split one such sub-latitude into two horizontal ones. The role of the $\beta$- and 
$\gamma$-points is to be the milestones on the path which indicate where it is possible
to make this splitting. And so, when the path meets a $\beta$-cline of type~2 along a leg, 
if the point of intersection is neither a $\beta$-point nor a $\gamma$-one, it knows that
it may cross this $\beta$-cline to go on the zig-zags. In the other case, depending on 
which type of point is met, the path knows that the $\beta$-cline must be followed in order
to cross the leg of a triangle.

\vskip 7pt
\noindent
$\underline{\hbox{From the generation~$n$ to the generation~$n$+1}}$
\vskip 7pt
   Figures~\ref{mauve_1} and~\ref{mauve_1in} allow us to prove the induction step
which allow to establish the path in a basic region of the generation~$n$+1 once
the path is established in any basic region of the generation~$n$.

   To see this point, it is enough to start from one of the figures~\ref{mauve_1}
or~\ref{mauve_1in}. Now, we consider that the big triangles belong to the generation~$n$+1
and that the small ones belong to the generation~$n$. Moreover, we assume that for
the generation~$n$, the path goes from an~$LP$ to the opposite~$HP$ inside a triangle
and from a~$HP$ to the opposite~$LP$ in the region in-between two consecutive triangles
of the generation~$n$ within the same latitude.

   It is not difficult to see that the same scheme of the indicated figures allow to
define the path for the basic regions of the generation~$n$+1. However, a tuning is needed
here, as the $\beta$-clines are no more in contact of the bases for the mauve triangles
of the generation~$n$+1. To see this point, consider that we also draw the mauve 
triangles of the generation~$n$$-$1, now assuming that~$n\geq1$. Then, it is not difficult to
see that the regions of the generation~$n$ split into regions of the generation~$n$$-$$1$ in
the same way as those of the generation~$n$+1 split into regions of the generation~$n$.

   We have the following general property:

\begin{lem}\label{cover}
Let~$\tau$ be a tile of the tiling. Then for any non-negative~$n$,
there is a mauve latitude $\Lambda$ of the generation~$n$ such
that $\tau\in\Lambda$. And then: either $\tau$ falls within a mauve
triangle of generation~$n$ in this latitude or $\tau$ falls outside
two consecutive mauve triangles of generation~$n$ and of the
latitude~$\Lambda$ and in between them.
\end{lem}

   This property follows immediately from the fact that the latitude of
a mauve triangle exactly covers that of the corresponding red triangle
and the following latitude of red phantoms.

\subsection{Additional tuning}

   In order to ensure the guidance of the path, we provide an additional tool.

   As indicated in the previous section, if one $LP$ allows the path to enter a triangle,
the other forbids such a possibility. We have the same property for a $HP$.

   In fact, it is not difficult to devise signals based on the notion of laterality
which allow to ensure this working. It may be one or the other $LP$, mandatory one of 
them and never both of them. This is performed by a signal which runs along the legs and
which meet at the vertex. Each $LP$ sends a signal to the other which runs along the leg to
the vertex where they meet. If the $LP$ admits the path, it sends a signal of its 
laterality and if not, it sends a signal of the other laterality. And so, it is
enough to forbid the meeting of signals of opposite lateralities. In this way, only
unilateral signals are allowed and they indicate the general motion of the path.
Note that once the laterality is fixed, this allows to place signboards at appropriate
places. First, the knowledge of which $LP$ is admitting allows to know which $HP$ allows
the path to exit from the triangle. This is inside a triangle. Now, the same mechanism
can be used to direct the path in between two consecutive triangles. This time, the 
information, still going from an~$LP$ to another goes through the corners and takes the
route of the red basis of a phantom which runs on the considered isocline. On this isocline
there can be corners of the appropriate generation only. Now, inside a basic region 
and within a sub-latitude, the direction of the path is the same. In fact it is the same
all along the latitude, as can be easily noticed from the fact that there is a shift
in the triangles with respect with the in between regions. Accordingly, the same 
direction occurs globally. The change of direction happens when the path meets the
legs of a triangle. This occurs for the standard hairpins of the zig-zags. Now, the
signal which goes from one $LP$ to the other also allows to place signboards at the
decisive positions: the mid-point and the $FP$'s, when the path climbs along the leg to
go from a sub-latitude to the next one. 
Now, the signal which goes in between two consecutive triangles
has also to detect the possibility of a leg coming from a bigger generation: this event
may change the direction of the further motion of the path. For this purpose, the
signal circumvents the mauve triangles it meets on its way. The isocline of a corner
continues a basis: accordingly it meets smaller mauve triangles at their $LP$'s,
the mauve triangles of their generation at corners again and bigger triangles at various
places, except the $LP$'s. Accordingly, when such a meeting occurs, the signal knows that
it stops here.
\vskip 5pt
   A last tuning deals with the parity of the number of zig-zags in a basic region.
  
   It is not difficult to notice that the path should arrive at particular isoclines
in the right direction. As already seen, the various signboards which we have constructed
allow to do this without problem. As an example, the corners of mauve triangles play an 
important role but there is not need to signalize them: they are recognizable by their
very conformation which is unique. Now, in order that the zig-zag line leads a point on 
a leg to the opposite leg, we need an odd number of zig-zags. The height of a triangle,
in terms of isoclines, the basis being included but the vertex being excluded, is an 
even number. But, it is not difficult to organize one piece of a zig-zag in a given
direction on two isoclines. It is enough to go up to a node of the highest isocline
from its leftmost son, then to go down to the next son from the son and then to go on until
the leftmost son of the next node. The need of such a run can be signalized, as the parity
of the number of isoclines can easily be computed. It is enough to put signboards of the 
required points three isoclines sooner in order the path know whether it goes on along 
a standard motion or it has to simultaneously cross two isoclines on the same motion. 
 
\section{About the injectivity of the global function of a cellular automaton in the
hyperbolic plane}

\subsection{Almost filling up the plane}

   We can derive two corollaries from lemma~\ref{cover}.

   The first one is very important:

\begin{cor}\label{nocycle}
The path contains no cycle.
\end{cor}

\noindent
Proof. I there was a cycle, it would be contained in some basic region. Now, the path
enters the region through an $LP$ in case of a triangle, through an $HP$ in case of an
in between region and it exits through the opposite~$HP$ or $LP$ respectively. Accordingly,
there is no cycle in this region due to the recursive structure of the basic regions
and it is plain that there is no cycle in a basic region of generation~0. \cqfd

Another one is also very important:

\begin{cor}\label{anysize}
For any tile~$\tau$, the path on one side of~$\tau$ fills up infinitely many mauve
triangles of increasing sizes.
\end{cor}

\noindent
Proof. This also comes from the filling up of the basic regions and their recursive
structure. The path cannot remain in the same latitude for ever. And so, it goes to another
latitude, upper or lower, depending on the structure of the underlying model implemented
by the interwoven triangles. This ensures the conclusion of the corollary. \cqfd

Another consequence is given by the following statement:

\begin{cor}\label{onecomponent}
If there are only finite basic regions, the path goes through any tile of the plane.
\end{cor}

\noindent
Proof. In this case, we have a sequence of increasing and embedded basic regions as
easily follows from the proof of corollaries~\ref{nocycle} and~\ref{anysize}.
\cqfd

\subsection{The exceptional situation}

   Corollary~\ref{onecomponent} indicates that if there are only finite triangles, then
we have a plane-filling path.

   Is it possible to have infinite triangles?

   The answer is yes: this means that there are also infinite red triangles. We know that
this happens with the butterfly model. No interwoven triangle crosses a given isocline~15.
As a corollary, there is an infinite mauve basis which crosses infinitely many 2-triangles
of any sizes. Now, this basis gives rise to infinitely many mauve triangles, by the very
principle of synchronization.
 
   And so, this situation is possible. Now, it is the unique one: an infinite triangle has an
infinite basis and this assumption leads to what we have just described.

   In this case, there cannot be a single path passing through all tiles of the plane once
only. 
  
   Indeed, once the path enters an infinite triangle, it cannot leave it. The same for 
a region in between two infinite triangles with the vertex on the infinite basis.
And so, in this case, there are infinitely many components for the path. However, 
corollary~\ref{anysize} is still valid for them.

\subsection{Proof of the main theorem}

   We can now prove:
   
\begin{thm}\label{undecinjfreeze}
The injectivity of the global function of a cellular automaton on the ternary heptagrid 
from its local transition function is undecidable.
\end{thm}

   The proof follows the argument of~\cite{jkari94}, with a slight modification. 

   In particular, we have to bring a new ingredient to the path as described in 
section~3: we define a direction for the path. This can be introduced by three hues
in the colour used for the signal of the path. One colour calls the next one and the last one calls the first one. The periodic repetition of this pattern together with the order
of the colours define the direction. This notion of direction allows to define 
the successor of a tile on the path. This can be formalized by a function $\delta$ 
from $Z\!\!\!Z$ to the tiling such that $\delta(n$+$1)$ is the successor of
$\delta(n)$ on the path.

   Remember that in~\cite{jkari94}, the automaton~$A_T$ attached to a set of 
tiles~$T$ has its states in $D\times\{0,1\}\times T$, where $D$ is the set of tiles 
which defines the tiling with the plane filling property and~$T$ is an arbitrary 
finite set of tiles. 
We can still tile the plane as we assume that the tiles of~$T$ are ternary heptagons
but the abutting conditions may be not observed: if it is observed with all the neighbours
of the cell~$c$, the corresponding configuration is said to be {\bf correct} at~$c$,
otherwise it is said {\bf incorrect}. When the considered configuration is correct at every
tile for~$D$ or at every tile for~$T$, it is called a {\bf realization} of the corresponding
tiling. Let~$\delta$ denote the function defining the orientation of the 
path induced by a realization of~$D$.  

   Here, we introduce a difference with~\cite{jkari94}. Instead of considering any
finite set of tiles, we consider the family $\{T_M\}$ of finite sets of prototiles defined
in~\cite{mmBEATCS} where $M$ runs over the set of deterministic Turing machines
with a single head and a single bi-infinite tape starting their computation from the empty
tape. As in~\cite{jkari94}, the transition function does not change neither the $D$- nor 
the \hbox{$T$-component} of the state of a cell~$c$: it only changes 
its $\{0,1\}$-component~$x(c)$ which is later on called the {\bf bit} of~$c$. As 
in~\cite{jkari94}, we define $A_{T_M}(x(c))=x(c)$ if the 
configuration in~$D$ or in~$T$ is incorrect at the considered tile. If both are correct, 
we define $A_{T_M}(x(c))=\hbox{\rm xor}(x(c),x(\delta(c)))$. It is plain that 
if $M$ does not halt, $T_M$ tiles the hyperbolic plane and there is a configuration 
of~$D$ and one of~$T_M$ which are realizations of the respective tilings. Then, the 
transition function 
computes the xor of the bit of a cell and its successor on the path. Hence, defining 
all cells with~0 and then all cells with~1 define two configurations which~$A_{T_M}$ 
transform to the same image: the configuration where all cells have the bit~0. 
Accordingly, $A_{T_M}$ is not injective.

Conversely, if $A_{T_M}$~is not injective, we have two different 
configurations~$c_0$ and~$c_1$ for which the image is the same. Hence, there is a 
cell~$t$ at which the configurations differ. Hence, the xor was applied, which means 
that~$D$ and~$T$ are both correct at this cell in these configurations and it is not 
difficult to see that the value for each configuration at the successor of~$t$ on
the path must also be different. And so, following the path in one direction, we have a 
correct tiling for both~$D$ and~$T_M$. Now, from corollary~\ref{anysize},
as the path fills up infinitely many triangles of increasing sizes, this
means that the tiling realized for $T_M$ is correct in these triangles. In particular,
the Turing machine~$M$ never halts. And so, we proved that $A_{T_M}$ is not injective
if and only if $M$ does not halt. Accordingly, the injectivity of $A_{T_M}$ is undecidable.
\cqfd

\section{Conclusion}

   The question of the surjectivity of the global function of cellular automata in the
hyperbolic plane is still open. In the Euclidean case, the undecidability of the
surjectivity problem is derived from the undecidability of the injectivity
as the surjectivity of the global function of a cellular automaton is equivalent to 
its injectivity on the set of finite configurations, see~\cite{moore,myhill}.
Now, in the case of cellular automata in the hyperbolic plane, this is not at all 
the case. The surjectivity and the injectivity of the global function are independent: 
there are examples of surjective global functions which are not injective and of 
injective global functions which are not surjective.

Accordingly, this question is completely open in the hyperbolic plane, even it
is is likely to be undecidable.

%

\end{document}